\documentclass[fleqn,usenatbib]{mnras}

\usepackage{newtxtext,newtxmath}
\usepackage{natbib}
\usepackage[T1]{fontenc}
\usepackage{ae,aecompl}
\usepackage{graphicx}	
\usepackage{amsmath}	
\usepackage{xcolor}
\usepackage[utf8]{inputenc}
\usepackage{float}
\usepackage{caption}
\usepackage{subcaption}
\usepackage{bm}
\usepackage{multirow}
\usepackage{placeins}
\usepackage{microtype}
\usepackage{graphicx} 
\usepackage{float}
\usepackage{comment}

\def\raa{RAA}

\title[Contribution of binary stars to EoR]{A CRASH simulation of the contribution of binary stars to the epoch of reionization}
\author[Ma et al.]{Qing-Bo Ma$^{1,2}$,\thanks{E-mail: \href{mailto:maqb@gznu.edu.cn}{maqb@gznu.edu.cn}}
Silvia Fiaschi$^{3}$,
Benedetta Ciardi$^{3}$,
Philipp Busch$^{4}$,
Marius B. Eide$^{3}$
\\
$^{1}$School of Physics and Electronic Science, Guizhou Normal University, Guiyang 550001, PR China \\
$^{2}$Guizhou Provincial Key Laboratory of Radio Astronomy and Data Processing, Guizhou Normal University, Guiyang 550001, PR China \\
$^{3}$Max-Planck-Institut f\"ur Astrophysik, Karl-Schwarzschild-Stra\ss e 1, D-85748 Garching bei M\"unchen, Germany\\
$^{4}$Department of Natural Science, The Open University of Israel, 1 University Road, P. O. Box 808, Raanana 43107, Israel
}

\date{Accepted XXX. Received YYY; in original form ZZZ}

\pubyear{2022}

\begin{document}
\label{firstpage}
\pagerange{\pageref{firstpage}--\pageref{lastpage}}
\maketitle

\begin{abstract}
We use a set of 3D radiative transfer simulations to study the effect that a large fraction of binary stars in galaxies during the epoch of reionization has on the physical properties of the intergalactic medium (i.e. the gas temperature and the ionization state of hydrogen and helium), on the topology of the ionized bubbles and on the 21 cm power spectra.
Consistently to previous literature, we find that the inclusion of binary stars can speed up the reionization process of HI and HeI, while HeII reionization is still dominated by more energetic sources, especially accreting black holes.
The earlier ionization attained with binary stars allows for more time for cooling and recombination, so that gas fully ionized by binary stars is typically colder than that ionized by single stars at any given redshift.  
With the same volume averaged ionization fraction, the inclusion of binary stars results in fewer small ionized bubbles and more large ones, with visible effects also on the large scales of the 21 cm power spectrum.  
\end{abstract}




\section{Introduction}
The epoch of reionization (EoR) starts after the first structures formed in the Universe \citep{Stark2016, Dayal2018}.
During the EoR, the cosmic gas drastically changes from being completely neutral to almost completely ionized and hot \citep{Furlanetto2006}. 
Some observations, e.g. of the spectra of high-$z$ quasars \citep[e.g.][]{Fan2006}, of the Cosmic Microwave Background (CMB) radiation \citep[e.g.][]{Planck2020}, and of the Ly$\alpha$ emitters \citep[e.g.][]{Weinberger2019}, suggest that the EoR is finished at $z>5$.
However, due to the abundant neutral hydrogen, the most promising probe to study EoR is the redshifted 21 cm line from neutral hydrogen \citep[e.g.][]{Field1959, Madau1997, Morales2010}, which can be measured by large arrays of low frequency radio telescopes, e.g. the Low-Frequency Array (LOFAR), the Square Kilometre Array (SKA), the Murchison Widefield Array (MWA), and the Hydrogen Epoch of Reionization Array (HERA).
Meanwhile, the most distant astronomical objects that belong to the EoR can be observed by e.g. the Hubble Space Telescope (HST; \citealt{Bouwens2015b}), the Atacama Large Millimeter/submillimeter Array (ALMA; e.g. \citealt{Wang2021}), and, in the near future, the James Webb Space Telescope (JWST; \citealt{Roberts2021}). 

The key topics in the theoretical studies of the EoR are \textit{how} this phase transformation occurred, and \textit{which} sources drove it. 
Previous studies in the literature have suggested that many sources can contribute to the reionization process, e.g. high-$z$ galaxies and quasars, X-ray binaries, and the supernova heated interstellar medium (ISM) \cite[see discussions in e.g. ][]{Madau1997, Loeb2001, Morales2010, Mesinger2013, Eide2018, Ross2019}.
However, it is still not clear in which proportion they do contribute.
To help answering this question, \citet[][hereafter E18]{Eide2018} and \citet[][hereafter E20]{Eide2020}  properly modeled UV and X-ray photon sources from stars in galaxies, accreting black holes, X-ray binaries (XRBs) and thermal bremsstralungh from the hot ISM, and investigated the effects of these possible source candidates on the ionization and heating of the IGM using sophisticated  multi-frequency radiative transfer (RT) simulations.
With the results of these simulations at hand, \citet{Ma2021} explored the associated  21 cm signal statistics (e.g. the global signal, power spectra and bispectra), while \citet{Ma2018} and \citet{Moriwaki2019} studied the cross-correlation of the 21 cm signal with the X-ray background and $\rm [O\, III]$ emitters, respectively.
Their results showed that stellar sources dominate hydrogen ionization, while gas heating is mostly provided by more energetic sources. These characteristics clearly appear in the 21 cm signal. 
However, one key factor which recently has attracted attention, i.e. the binary star interaction \cite[e.g. ][]{MaX2016, Stanway2016, Rosdahl2018, Gotberg2020, Doughty2021}, has not been taken into account in their calculations.

Observations of the local Universe show that at least 50\% of stars are in binary systems \citep{Han2020}.
The interaction of binaries can alter the evolutionary tracks of their constituents by stripping the envelope of massive stars, resulting in higher UV emissivities and harder spectra emitted over a prolonged lifetime \citep{Eldridge2012, Gotberg2019, Berzin2021}, with photons hard enough to be able to doubly ionize helium \citep{Gotberg2020}.
Indeed, results of stellar population synthesis codes \citep{Stanway2016, Gotberg2020} show that the inclusion of binary systems produces an ionizing flux which is e.g. 60\% higher than single stars in populations with metallicity $0.05 Z_{\odot} \leq Z \leq 0.3 Z_{\odot}$ and 10-20\% in populations with near-solar metallicities.
Binary stars are also at the origin of e.g. X-ray binaries, supernovae and blue stragglers \citep{Han2020}.
Additionally, several works showed that the inclusion of binary stars can increase the escape fraction of ionizing photons from galaxies \cite[e.g. ][]{MaX2016, MaX2020, Secunda2020}, as their stronger stellar feedback can clear the surrounding gas clouds.
If a large fraction of stars during the EoR is then formed in binary systems, we can expect that the combination of the two effects mentioned above could substantially increase the budget of photons available to ionize the IGM, speeding up the reionization process (see e.g. \citealt{Rosdahl2018}). 
The results of hydrodynamic cosmological simulations including RT processes \citep{Doughty2021} confirmed that a high binary fraction in galaxies results in a faster reionization of HI, although at the same time the star formation rate of low-mass galaxies is reduced due to the increased heating.

In this paper, we use high-resolution hydro-dynamical and RT simulations to study the effects of binary stars on the IGM properties, i.e. HI, HeI and HeII ionization and gas temperature. 
To do so, we extend the work of E18 and E20 by including the contribution of binary stars.
Differently from previous studies \cite[e.g. ][]{Rosdahl2018, Doughty2021}, we will focus on the differences that binary stars induce on the IGM properties in comparison to single stars, also in the presence of energetic sources. 
We will additionally evaluate their effects on the 21 cm power spectra and the topology of ionized regions.

The rest of paper is organized as follows: the simulations and models adopted are described in Section~\ref{Sec2:method}; Section~\ref{Sec3:res} presents the results of the binary star effects on the IGM properties, 21 cm signal, and the topology of ionized bubbles, while the conclusions are summarized in Section~\ref{Sec4:con}.

\section{Method}
\label{Sec2:method}
To model the process of reionization, we follow the approach presented in E18 and E20, which combines the outputs of high-resolution cosmological hydrodynamical simulations (subsection \ref{MBII}) with radiative transfer (subsection \ref{CRASH}). 
In this section we highlight the main aspects and refer the reader to the original papers for more details. 

\subsection{Cosmological hydrodynamic simulation} 
\label{MBII}
We employ the results of the hydrodynamical cosmological simulation MassiveBlack-II (MB-II,  \citealt{Khandai2015}), a state-of-the-art high resolution and large volume SPH simulation, which has been run with \textsc{P-GADGET}, a modified version of \textsc{GADGET-3} (see \citealt{Springel2005} for details about the earlier version).
It has a box length of $100 \, h^{-1}\, \mathrm{cMpc}$, with the cosmological parameters \citep{Komatsu2011}: $\sigma_8=0.816$,  $n_s=0.968$, $\Omega_{\Lambda}=0.725$, $\Omega_{m}=0.275$, $\Omega_{b}=0.046$ , and $h=0.701$. 
The total number of dark matter and gas particles is 2$\times$1792$^3$, with mass $m_{\rm DM}=1.1\times10^7 \, h^{-1} \, \mathrm{M_{\odot}}$ and $m_{\rm gas}=2.2\times10^6 \, h^{-1} \, \mathrm{M_{\odot}}$, respectively.

In addition to the baryonic physics and feedback processes \citep{Di2008, Croft2009, Degraf2010, Di2012}, MB-II includes a multi-phase interstellar medium model for star formation \citep{Springel2005}, and tracks stellar populations, galaxies, accreting and dormant black holes, as well as their properties (e.g. position, age, metallicity, mass, accretion rate and star formation rate). 

15 MB-II snapshots are employed to cover the evolution between $z = 20$ and $5$. 
The particles of each snapshot are mapped onto a Cartesian grid of $256^3$ cells (corresponding to a spatial resolution of $391 \, h^{-1} \, \mathrm{cKpc}$) to create maps of gas mass density and initial gas temperature.
The gas density is then converted to hydrogen and helium number densities by assuming a mass fraction of $X=0.76$ and $Y=0.24$, respectively.
Note that the metallicity is ignored in this case, as we are concentrating on the overwhelming volume contribution of pristine IGM, which is far outside the potentially contaminated circumgalactic medium. 

\subsection{Radiative transfer simulation}
\label{CRASH}
With the information about the properties of galaxies from MB-II, four types $s$ of ionizing and heating sources are modeled (see E18 and E20):
\begin{itemize}
    \item single and/or binary stars (hereafter abbreviated as SS and BS, respectively), whose spectra and luminosities are evaluated from the age, metallicity and mass of stellar particles using the stellar population synthesis code Binary Population and Spectral Synthesis (BPASS; version 1.1, \citealt{Eldridge2012});
    \item neutron star/black hole X-ray binaries, hereafter XRBs, whose spectra and luminosities depend on the star formation rate and the metallicity of gas particles \citep{Madau2017};
    \item thermal bremsstrahlung from supernova-heated interstellar medium, hereafter ISM, which is modelled as a broken power-law spectrum with a luminosity proportional to the star formation rate of the host galaxy \citep{Mineo2012};
    \item  accreting nuclear black holes, hereafter BHs, whose luminosity is proportional to their accretion rate, while an average spectrum is adopted based on quasars' observations at lower redshift \citep{Krawczyk2013}.
\end{itemize}
The Spectral Energy Distribution (SED) and luminosity of each source $i$ of type $s$ are indicated as $S_i^s$ (erg~s$^{-1}$~Hz$^{-1}$) and $L_i^s$ (erg~s$^{-1}$), respectively. 
These sources are also mapped onto the $256^3$ cells.
If different source types happen to be in the same cell, their properties are summed up to obtain $S_i=\sum_s S_i^s$ and $L_i=\sum_s L_i^s$, except for the BHs which are added as separate sources at the same coordinates.
Although each single source could be assigned a different spectrum, to simplify the overall procedure, an averaged SED, $\bar{S}=\langle S_i \rangle$, is adopted at each $z$. 
The BHs spectra are calculated separately as $\bar{S}^{\mathrm{BH}}= \langle S^{\mathrm{BH}}_i \rangle$. 

The radiative transfer simulations are performed by post-processing the outputs of MB-II with the multi-frequency Monte Carlo ray-tracing code \textsc{CRASH} \citep{Ciardi2001, Maselli2003, Graziani2013, Hariharan2017,  Graziani2018, Glatzle2019}, which produces the ionization state of hydrogen and helium, as well as the gas temperature. 
We highlight that each source spectrum in CRASH is discretized into 82 frequency bins with $h_{\rm p}\nu \in [13.6\,\mathrm{eV}, 2\,\mathrm{KeV}]$, where $h_{\rm p}$ is the Planck constant. 
The bins are spaced more densely around the ionization thresholds of hydrogen (13.6 eV) and helium (24.6 eV and 54.4 eV).

While most inputs for the radiative transfer simulations are directly derived from physical properties predicted by the hydrodynamic simulations, the only parameter which needs to be fixed is   the escape fraction of ionizing photons $f_{\rm esc}$, which we apply to  radiation below 200 eV.
Recent studies have claimed that $f_{\rm esc}$ may have evolved with redshift $z$, as galactic properties changed (e.g. \citealt{Paardekooper2015, Trebitsch2017, Katz2018}). 
Thus, the $f_{\rm esc}$ adopted here is (see also E20 and \citealt{Price2016}):
\begin{equation}
    f_{\mathrm{esc}}(z)=0.15 \times \left ( \frac{1+z}{9} \right )^{\beta},
\end{equation}
where $\beta=2.2$, so that at high-$z$ $f_{\rm esc} \simeq 100\%$ while it becomes $8.6\%$ at $z=6$. 
This choice yields emissivities consistent with the measurements of \cite{Bouwens2015}.

In our analysis we include four simulations. 
Two of them are from E18 and E20, i.e. the one including only SS in galaxies (named $\mathrm{STARS^{SS}}$) and the other one including SS, ISM, BHs and XRBs (named $\mathrm{ALL^{SS}}$).
We additionally run two new simulations, one includes BS in galaxies (named $\mathrm{STARS^{BS}}$), and one includes the contribution from BS, ISM, BHs and XRBs (named $\mathrm{ALL^{BS}}$).

\begin{figure}
\centering
	\includegraphics[width=0.96\linewidth]{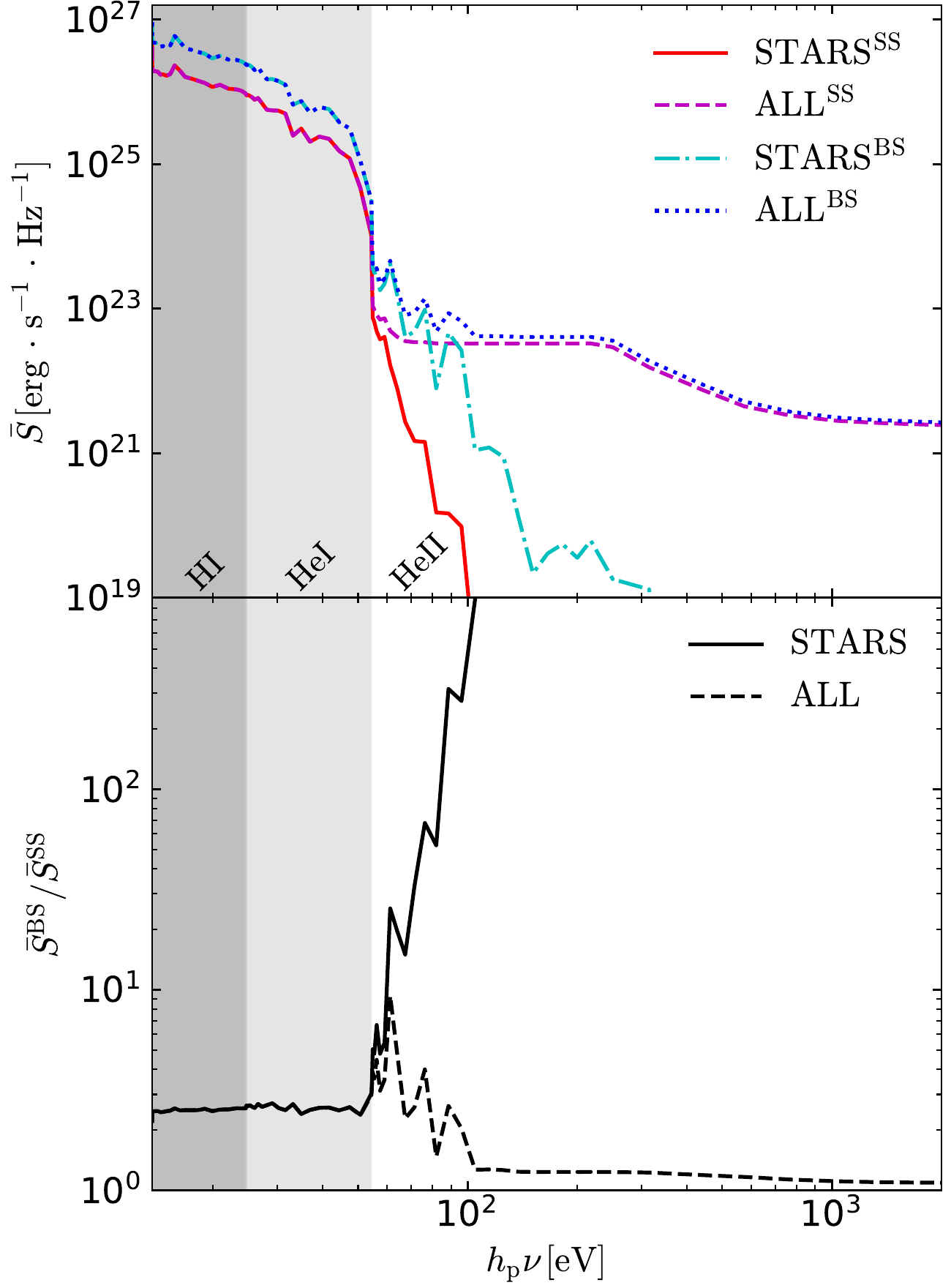}
    \caption{{\it Upper panel}: Globally averaged spectral energy distribution, $\bar{S}$, at $z=7.5$ for simulation $\mathrm{STARS^{SS}}$ (red solid line), $\mathrm{ALL^{SS}}$ (magenta dashed line), $\mathrm{STARS^{BS}}$ (cyan dash-dotted line) and $\mathrm{ALL^{BS}}$ (blue dotted line).  
    The vertical gray regions indicate the ionization thresholds for hydrogen ($\ge$ 13.6 eV), neutral helium ($\ge$24.6 eV) and singly ionized helium ($\ge$54.4 eV). 
    {\it Lower panel}: SED ratios for simulations with/without binary stars ($\bar{S}^{\rm BS} / \bar{S}^{\rm SS}$), with only stellar type sources (solid line) and including also the more energetic sources (dashed line).
    Note that the SED of BHs is not included.
    }
    \label{fig:SED}
\end{figure}
As a reference, in Fig.~\ref{fig:SED} we plot $\bar{S}$ for the four simulations at $z=7.5$. 
At $h_{\rm p}\nu \lesssim$ 60 eV, the SEDs are dominated by the stars (indeed in this range the SEDs for both STARS and ALL simulations are virtually the same), while the contribution of energetic sources, i.e. ISM and XRB, becomes significant at higher energies (hard UV and soft X-ray), those above the ionization threshold of singly ionized helium (HeII). 
The presence of BS extends the dominance of the stellar SED up to $\sim 100$~eV.
Meanwhile, it increases the SED by up to 150\% and 160\% in the (13.6--24.6)~eV and  (24.6--54.4)~eV bands, respectively. 
Without energetic sources, i.e. the STARS simulations, their contribution to the SED above the HeII ionization threshold ($>$ 54.4~eV) is even larger (see $\bar{S}^{\rm BS} / \bar{S}^{\rm SS}$ values in the bottom panel of the figure). 
When all sources are included, the difference between the BS and SS case is only clearly visible at $h_{\rm p}\nu \lesssim$ 100 eV, while at larger energies $\bar{S}^{\rm BS} / \bar{S}^{\rm SS} \approx 1$.

\begin{figure}
\centering
	\includegraphics[width=0.92\linewidth]{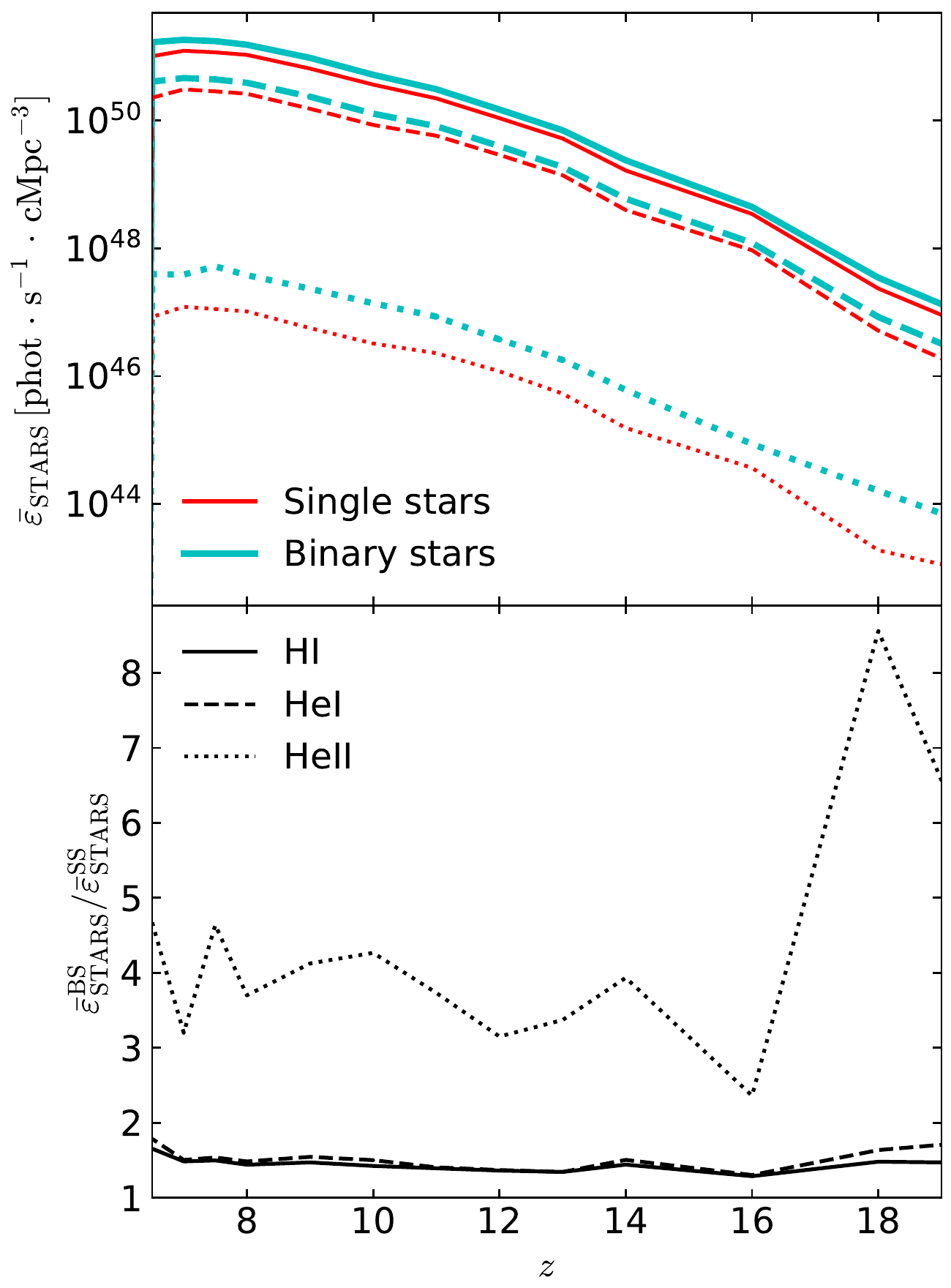}
    \caption{{\it Upper panel}:
    Redshift evolution of the comoving volume averaged emissivity $\bar{\varepsilon}_{\rm STARS}$ of HI (solid lines), HeI (dashed lines) and HeII (dotted lines) ionizing photons from single stars (SS, red thin lines) and binary stars (BS, cyan thick lines), respectively.
    {\it Lower panel}: Redshift evolution of $\bar{\varepsilon}_{\rm STARS}^{\rm BS} / \bar{\varepsilon}_{\rm STARS}^{\rm SS}$ for HI (solid line), HeI (dashed line) and HeII (dotted line).
    }
    \label{fig:Emissivity_species}
\end{figure}
The top panel of Fig.~\ref{fig:Emissivity_species} shows the volume averaged emissivity  $\bar{\varepsilon}_{\rm STARS}$ of HI, HeI and HeII ionizing photons, i.e. with energy $h_{\rm p}\nu>13.6$, 24.6 and 54.4~eV, respectively, from SS (i.e. simulation $\mathrm{STARS^{SS}}$) and BS (i.e. simulation $\mathrm{STARS^{BS}}$).
Note that because the contribution to the ionization budget from energetic sources, i.e. ISM, XRB and BH, is negligible (see E18 and E20), we do not plot the results of simulations $\mathrm{ALL^{SS}}$ and $\mathrm{ALL^{BS}}$.
As expected, both BS and SS have very high emissivity of HI ionizing photons, which increases quickly with decreasing redshift.
In both simulations, the emissivity of HeI ionizing photons is $\sim 25\%$ of the HI one, while the HeI number density is only $\sim 7.9\%$ of HI. 
Thus, even accounting for the higher helium recombination rate, we expect the ionization of HeI to be slightly higher than that of HI, due also to the larger photoionization cross section of the former. 
The emissivity of HeII ionizing photons is only $0.1\%$ of that for HeI ionizing photons, so we do not expect very high HeIII fractions in either simulation.
BS produce an ionizing photon emissivity higher than that of SS. 
As a reference, at $z=7.5$ the volume averaged emissivity for HI, HeI and HeII ionizing photons is
$1.75\times10^{51}$ ($1.17\times10^{51}$), $4.42\times10^{50}$ ($2.87\times10^{50}$) and   $5.21\times10^{47}$ ($1.12\times10^{47}$)  phot~s$^{-1}$~cMpc$^{-3}$, respectively for BS (SS). 
In the bottom panel of Fig.~\ref{fig:Emissivity_species}, we show the ratio between $\bar{\varepsilon}_{\rm STARS}^{\rm BS}$ and $\bar{\varepsilon}_{\rm STARS}^{\rm SS}$, which is $\sim$1.5 for HI and HeI ionizing photons in the whole redshift range, while it is $\sim 3-5$ (with the exception of very high redshift when it is much higher because of the lower metallicity\footnote{Note that at $z=20$ and 18 there are only 1 and 26 sources, respectively, so that the scatter is very large.}) for HeII ionizing photons, due to the harder spectrum of BS compared to SS. 

Finally, we note that the version of BPASS used in this work is the same of E18 and E20 (v1.1), and not the most recent one (v2.2.1; \citealt{Stanway2018}), to make a consistent comparison with simulations including only single stars, and avoid differences of the spectral synthesis code introduced by the 2018 and 2012 versions.
The v2.2.1 produces harder spectra, due to the improved prescription for the rejuvenation of secondary stars and the inclusion of an extended and finely sampled grid of low mass stellar evolution models. 

\section{Results}
\label{Sec3:res}
In the following, we will discuss results from the two simulations we have run, i.e. STARS$^{\rm BS}$ and ALL$^{\rm BS}$, which will also be compared to the equivalent simulations from E18 and E20 without the contribution from BS (i.e. STARS$^{\rm SS}$ and ALL$^{\rm SS}$) in terms of IGM properties (subsections \ref{snapshot} and \ref{IGM_prop}), 21 cm signal (subsection \ref{21cm_sig}), and topology of ionized bubbles (subsection \ref{sec:topology}). 

\subsection{Qualitative overview of IGM properties}
\label{snapshot}
\begin{figure*}
\centering
	\includegraphics[width=0.96\linewidth]{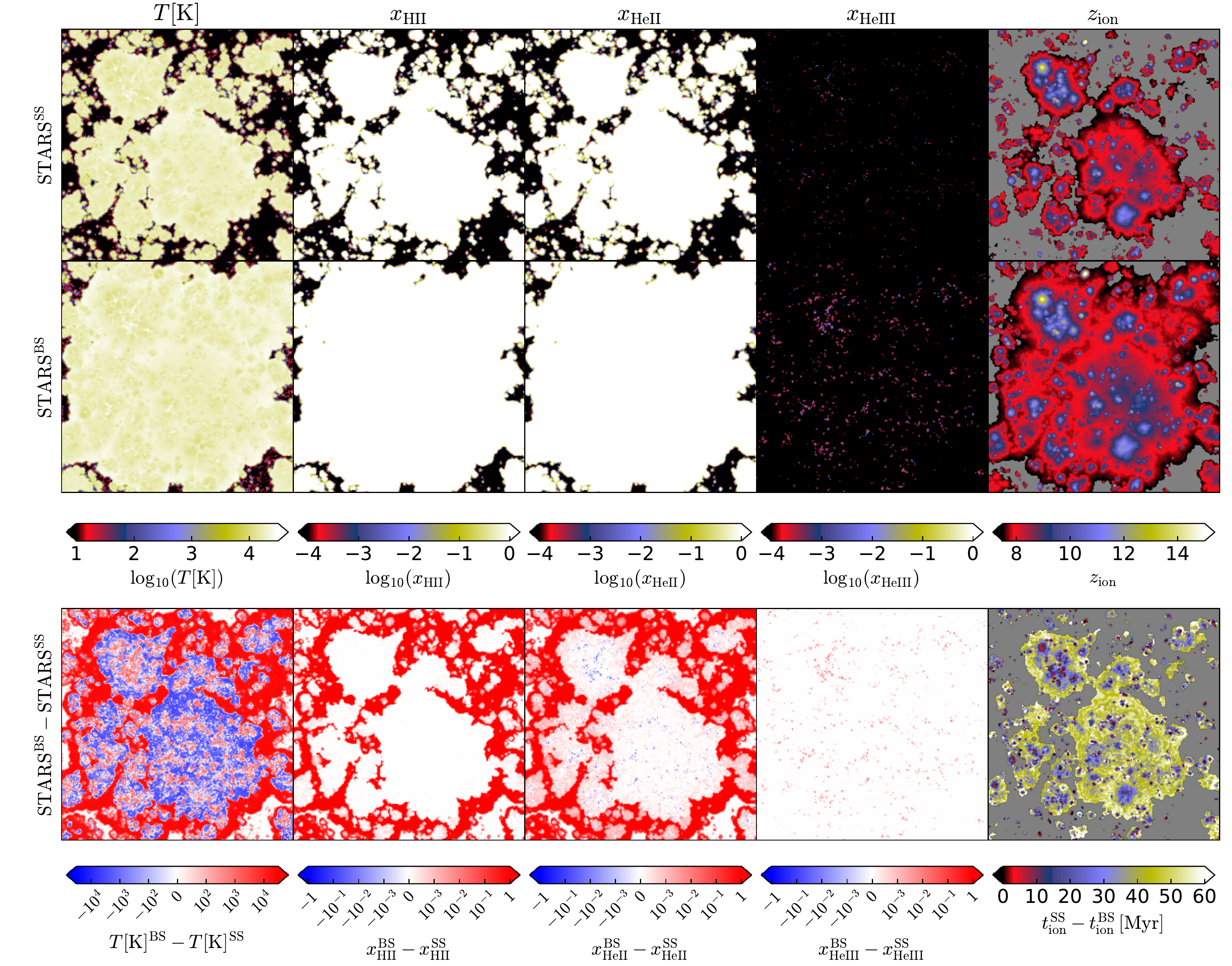}
    \caption{From left to right, maps of gas temperature ($T$), HII ($x_{\rm HII}$), HeII ($x_{\rm HeII}$) and HeIII ($x_{\rm HeIII}$) fractions, and redshifts of cells ionized for the first time ($z_{\rm ion}$) of central slices through the simulation volume at $z$ = 7.5 for the simulations $\mathrm{STARS^{SS}}$ (top) and $\mathrm{STARS^{BS}}$ (middle). 
    The bottom maps correspond to the difference between the two simulations, i.e. $\mathrm{STARS^{BS} - STARS^{SS}}$.
    The last map shows the differences of cosmic time corresponding to $z_{\rm ion}$. 
    The maps correspond to the full box length of  $100\ h^{-1}$ cMpc, and have a thickness of $0.39\ h^{-1}$ cMpc, i.e. one cell.
    Note that the values of $x_{\rm HII}$, $x_{\rm HeII}$, and $x_{\rm HeIII}$ are truncated below $10^{-4}$, and the grey regions in the $z_{\rm ion}$ and time difference maps denotes the cells still neutral at $z=7.5$.
    }
    \label{fig:GAL_z_7.5}
\end{figure*}
Fig.~\ref{fig:GAL_z_7.5} displays map samples of the thermal and ionization state of the IGM at $z=7.5$ when only stellar sources are included in the simulations, both for the cases with SS (i.e. $\mathrm{STARS^{SS}}$) and with BS (i.e. $\mathrm{STARS^{BS}}$), together with their difference maps. 
As a reference, we also show the redshifts of cells ionized for the first time, $z_{\rm ion}$, at $z>7.5$, and the corresponding cosmic time difference between cases with SS and BS.
Without the heating of energetic sources, the gas is basically either neutral and cold (with $T < 10\, \rm K$ and $x_{\rm HII} \approx x_{\rm HeII} \approx x_{\rm HeIII} \lesssim 10^{-4}$) or hot and fully ionized (i.e. $T >10^{4}\,\rm K$ and $x_{\rm HII} \approx x_{\rm HeII} \sim 1$).
Due to the larger number of ionizing photons produced by BS, the extent of the ionized regions is larger for all components, i.e. HII, HeII and HeIII. 
In addition, because of the harder spectrum produced by BS, ionization of He is more effective, so that actual HeIII regions are now clearly visible, although they are still very small.
We note that the predominance of ionization towards the center of the maps is due both to the density and source distribution, but also to the lack of periodic boundary conditions in the radiative transfer simulations (see discussion in the Appendix of E20 for more details). 

The differences caused by the inclusion of BS are better appreciated in the lower panels of Fig.~\ref{fig:GAL_z_7.5}. 
The larger extent of the ionized regions obtained in the presence of BS results in the dark red cells in the $T$, $x_{\rm HII}$ and $x_{\rm HeII}$ maps, denoting gas which has been fully ionized in $\mathrm{STARS^{BS}}$ but not in $\mathrm{STARS^{SS}}$, and indeed such gas has $T>10^{4}\,\rm K$.
Cells which are neutral in both simulations have similar $T$, $x_{\rm HII}$ and $x_{\rm HeII}$, with differences close to zero. This indicates that BS have no obvious effect on the properties of the neutral gas surrounding the ionized regions.

It is particularly interesting to note that, while for the HII component the difference is limited to a larger extent of the ionized regions in the presence of BS, this is not the case for the HeII regions and temperature, which show a more complex structure. Indeed, a slight excess of HeII is observed due to their harder spectra, which is also responsible for the full ionization of He in some pockets of gas (red cells in the HeIII map), resulting in a depletion of HeII (blue cells in the HeII map).
Gas with fully ionized helium (in general cells in the vicinity of the sources) is also hotter (red cells within the fully ionized regions in the temperature map), while cells further away from the sources are ionized earlier in $\mathrm{STARS^{BS}}$ than in $\mathrm{STARS^{SS}}$ (see the last column in Fig.~\ref{fig:GAL_z_7.5}), and thus have more time to cool down (appearing blue in the difference map).
In fact, the mean temperature of the ionized cells in simulation $\mathrm{STARS^{SS}}$ is $1.77\times 10^{4}\,\rm K$, while the one of the same (ionized) cells in $\mathrm{STARS^{BS}}$ is $1.72\times 10^{4}\,\rm K$.
However, the mean temperature of all ionized cells in $\mathrm{STARS^{BS}}$ is $1.88\times 10^{4}\,\rm K$, i.e. gas freshly ionized by BS is hotter than that was ionized by SS. 

\begin{figure*}
	\includegraphics[width=0.96\linewidth]{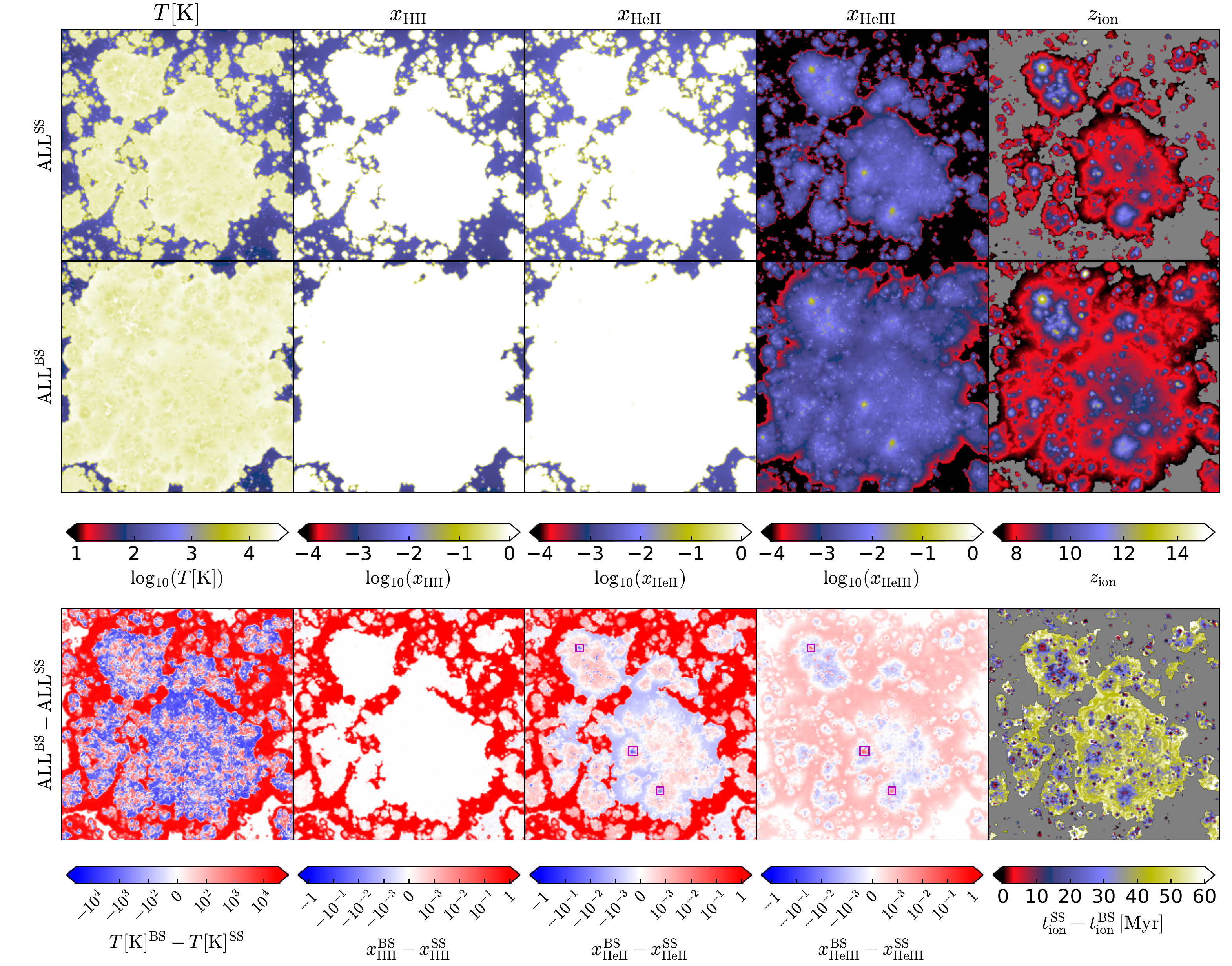}
    \caption{Same as Fig.~\ref{fig:GAL_z_7.5} but for $\mathrm{ALL^{SS}}$ (top), $\mathrm{ALL^{BS}}$ (middle), and their differences $\mathrm{ALL^{BS} - ALL^{SS}}$ (bottom). The magenta square boxes  denote three representative locations where the HeIII production is enhanced by the presence of binary stars. }
    \label{fig:ALL_z_7.5}
\end{figure*}
In Fig.~\ref{fig:ALL_z_7.5} we show instead the impact of binary stars when also sources other than stars are included, i.e. XRBs, ISM and BHs (simulations with ALL).
While we refer the reader to E18 and E20 for a more extended discussion, here we just highlight that, although the extent and characteristics of the fully ionized regions is dictated by stars (as can be seen from a comparison with Fig.~\ref{fig:GAL_z_7.5}), more energetic sources produce partial ionization and heating outside of the fully ionized gas due to the longer mean free path of their photons, as well as larger HeIII fractions. The production of the latter component is dominated by BHs, as highlighted in E18 and E20. Here though, we concentrate our discussion on the effect of BS.

The impact of BS on $T$ and $x_{\rm HII}$ is alike the one observed in the STARS simulations, and indeed the difference maps are very similar to those in Fig.~\ref{fig:GAL_z_7.5}.
As a reference, the mean temperature of the regions ionized in both simulations, $\mathrm{ALL^{SS}}$ and $\mathrm{ALL^{BS}}$, is $1.79\times 10^{4}\,\rm K$ and $1.75\times 10^{4}\,\rm K$, respectively, while the one of all ionized cells in simulation $\mathrm{ALL^{BS}}$ is $1.90\times 10^{4}\,\rm K$. 
The HeII and HeIII maps, though, show some differences. 
As already seen in E20, in the presence of energetic sources, within HII (and HeII) regions $x_{\rm HeIII}$ can reach values $\gtrsim 10^{-3}$.
As a comparison, the $x_{\rm HeIII}$ of neutral cells in both simulations is $<10^{-4}$.
As observed also in the STARS simulations, in the vicinity of the sources the production of HeIII is enhanced by the presence of BS (i.e. concentrated red points in the HeIII difference map, some of which are flagged by magenta square boxes), resulting in a lower value of $x_{\rm HeII}$ (i.e. concentrated blue points in the HeII difference map, see the corresponding magenta square boxes at the same positions). 
Further away from the sources the details of the ionization state of the gas are very sensitive both to the recombination/ionization balance and the timing of reionization, which are roughly consistent with the cosmic time differences in the last column of Fig.~\ref{fig:ALL_z_7.5}. 
The earlier strong HeII ionization (with $x_{\rm HeIII} > 10^{-2}$) experienced by some cells in the presence of BS is subsequently not sustained against their high recombination rate, so that at $z=7.5$ they show a $x_{\rm HeIII}$ lower than $\mathrm{ALL^{SS}}$ (blue cells in the HeIII difference map), and a correspondingly higher $x_{\rm HeII}$. 
This trend is reversed for more freshly ionized cells, i.e. further away from the sources. The reason for this is that, although HeII is ionized earlier in $\mathrm{ALL^{BS}}$ than in $\mathrm{ALL^{SS}}$, here $x_{\rm HeIII}$ is small ($\sim 10^{-3}$), with a comparatively low recombination rate, so that in this case the longer time of shining by energetic sources leads to higher $x_{\rm HeIII}$ in $\mathrm{ALL^{BS}}$.

\subsection{Evolution of IGM properties}
\label{IGM_prop}
\begin{figure*}
\centering
	\includegraphics[width=0.92\linewidth]{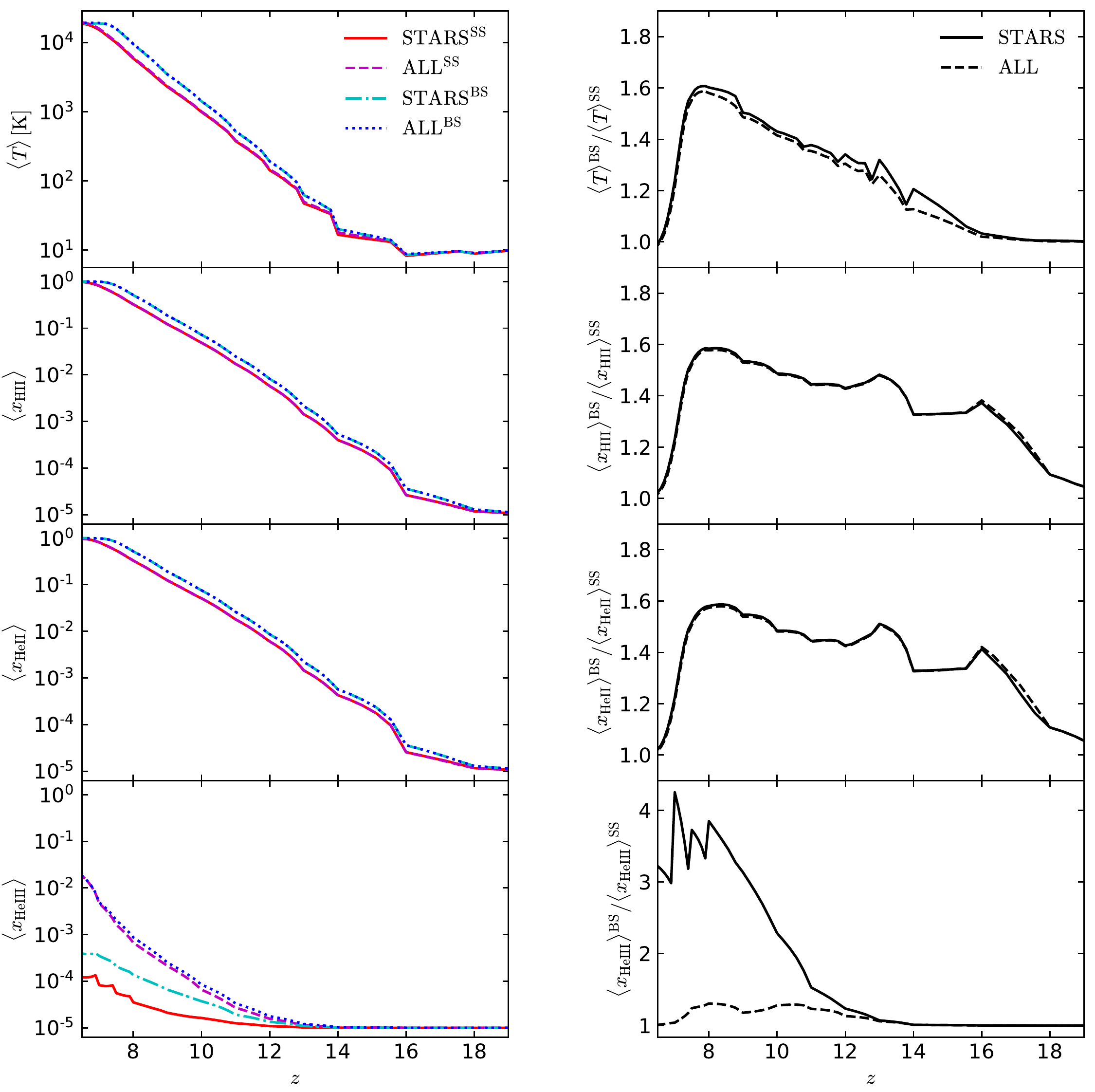}
    \caption{{\it Left panels:} Evolution of the volume averaged (from top to bottom) gas temperature $\langle T \rangle$, and ionization fraction $\langle x_{\rm HII}\rangle$, $\langle x_{\rm HeII} \rangle$ and $\langle x_{\rm HeIII} \rangle$, for simulations $\mathrm{STARS^{SS}}$ (red solid lines),  $\mathrm{ALL^{SS}}$ (magenta dashed), $\mathrm{STARS^{BS}}$ (cyan dash-dotted) and $\mathrm{ALL^{BS}}$ (blue dotted). 
    {\it Right panels:} relative ratio between BS and SS for simulations with STARS (solid) and ALL (dashed).
    }
    \label{fig:averaged}
\end{figure*} 
Fig.~\ref{fig:averaged} displays the redshift evolution of the volume averaged IGM temperature $\langle T \rangle$, and ionization fractions $\langle x_{\rm HII} \rangle$, $\langle x_{\rm HeII} \rangle$, and $\langle x_{\rm HeIII} \rangle$ in our four simulations, together with the ratio between results of simulations with BS and SS. 
\begin{table}
\scriptsize{}
\setlength\arrayrulewidth{0.5pt}
\begin{tabular}{lcccccc}
\hline
\textbf{\small Simulation} &  & \textbf{\small $\langle T \rangle\, [{\rm K}]$} & \textbf{\small{}$\left\langle x_{\mathrm{HII}}\right\rangle $} & \textbf{\small $\left\langle x_{\mathrm{HeII}}\right\rangle $} & \textbf{\small $\left\langle x_{\mathrm{HeIII}}\right\rangle $} \tabularnewline
\hline 
\textbf{\footnotesize{}$z=9$} &  &  &  &  & \tabularnewline
\hline 
$\mathrm{STARS^{SS}}$ &  & $2288$ & $0.1209$ &  $0.1232$ & $2.104\times10^{-5}$\tabularnewline
$\mathrm{ALL^{SS}}$ &  & $2338$ &  $0.1219$ &  $0.1244$ &  $2.134\times10^{-4}$\tabularnewline
$\mathrm{STARS^{BS}}$ &  & $3439$ & $0.1855$ &  $0.1905$ & $6.595\times10^{-5}$\tabularnewline
$\mathrm{ALL^{BS}}$ &  & $3474$ & $0.1865$ &  $0.1914$ &  $2.513\times10^{-4}$\tabularnewline
\hline 
\textbf{\footnotesize{}$z=7.5$} &  &  &  &  & \tabularnewline
\hline 
$\mathrm{STARS^{SS}}$ &  &  $9935$ &  $0.5338$ &  $0.5418$ &  $5.563\times10^{-5}$\tabularnewline
$\mathrm{ALL^{SS}}$ &  & $10222$ &  $0.5403$ &  $0.5473$ &  $1.625\times10^{-3}$\tabularnewline
$\mathrm{STARS^{BS}}$ &  & $15633$ &  $0.8169$ &  $0.8216$ &  $2.073\times10^{-4}$\tabularnewline
$\mathrm{ALL^{BS}}$ &  & $15871$ & $0.8215$ &  $0.8245$ &  $2.017\times10^{-3}$\tabularnewline
\hline 
\textbf{\footnotesize{}$z=6.5$} &  &  &  &  & \tabularnewline
\hline 
$\mathrm{STARS^{SS}}$ &  &  $18706$ & $0.9783$ & $0.9789$ &  $1.212\times10^{-4}$\tabularnewline
$\mathrm{ALL^{SS}}$ &  & $19437$ & $0.983$ &  $0.9655$ & $1.805\times10^{-2}$\tabularnewline
$\mathrm{STARS^{BS}}$ &  & $18570$ & $0.9999$ & $0.9995$ & $3.904\times10^{-4}$\tabularnewline
$\mathrm{ALL^{BS}}$ &  & $19185$ & $0.9999$ &  $0.9817$ & $1.818\times10^{-2}$\tabularnewline
\hline 
\end{tabular}
\caption{Volume averaged gas temperature $\langle T \rangle$ and ionization fractions $\langle x_{\rm HII} \rangle$, $\langle x_{\rm HeII} \rangle$ and $\langle x_{\rm HeIII} \rangle$ at $z$ = 9, 7.5 and 6.5 for our four simulations.
}
\label{table_values_z}
\end{table}
As a reference, Table~\ref{table_values_z} lists these values at $z=9$, 7.5 and 6.5.
Similar plots with the evolution as a function of HII fraction $\langle x_{\rm HII} \rangle$ are shown in Fig.~\ref{fig:averaged_vi} of appendix \ref{appendix:pro_vi}.

During the early stages of reionization, when most of the IGM is still cold and neutral, $\langle T \rangle$ is dominated by the hydrodynamical temperature from the MB-II, and indeed the four simulations show very similar values.
$\langle T \rangle$ rises up quickly with decreasing redshift, with values from simulations including BS  becoming increasingly larger than those with SS, as can be clearly seen by the ratios between BS and SS simulations shown in the right hand side of the figure (it reaches a maximum of $\sim 1.6$ at $z \approx 8$). 
The ratio is slightly lower in the ALL than the STARS simulations because the heating effect of BS is dampened by the presence of more energetic sources. 
Once (almost) full ionization is reached in both simulations, $\langle T \rangle$ converges again to similar values (see also Table~\ref{table_values_z}). 
It is interesting to note that by $z \approx 6.5$ the mean temperature $\langle T \rangle$ in simulations with BS is slightly lower than in those with SS (see also Table~\ref{table_values_z}). 
This is again related to the earlier ionization reached in the presence of BS, as already discussed in E20 (and previously in \citealt{Keating2018}) recently ionized gas is hotter than earlier ionized one. 
As also shown in E18 and E20, $\langle T \rangle$ is only slightly increased by the presence of more energetic sources (see also Table~\ref{table_values_z}).

As discussed extensively in E18 and E20 (see also Table~\ref{table_values_z}), the energetic sources negligibly increase the averaged fractions of HII and HeII.
Conversely, simulations including BS result in visibly larger ionization fractions at any given redshift (as also \citealt{MaX2016} and \citealt{Gotberg2020}), e.g. with $\langle x_{\rm HII} \rangle$ and $\langle x_{\rm HeII} \rangle$ more than $60\%$ higher than those obtained in simulations with SS at $z \sim 8$.
Their differences become much smaller ($\sim$ a few percent) towards the end of the EoR, when $\mathrm{\langle x_{HII} \rangle \cong \langle x_{HeII} \rangle} \sim 1$. 
In all four simulations, initially $\langle x_{\rm HeII} \rangle$ is slightly higher than  $\langle x_{\rm HII} \rangle$ (see e.g. the values at $z = 9$ and 7.5 in Table~\ref{table_values_z}) due to the larger emissivity of HeI ionizing photons compared to its number density (see discussion in subsection~\ref{CRASH}) and also to the larger cross-section of HeI to UV photons. 
At later times, instead, when the Universe is highly ionized (see e.g. at $z = 6.5$ in Table~\ref{table_values_z}), $\langle x_{\rm HII} \rangle$ becomes larger than $\langle x_{\rm HeII} \rangle$, since the stronger hard UV and X-ray radiation from BS and energetic sources produce a non-negligible amount of HeIII, depleting HeII. 
The only exception is $\mathrm{STARS^{SS}}$, as SS are not able to produce a significant amount of HeIII.
Indeed, simulations including BS and/or energetic sources all reach values of $\langle x_{\rm HeIII} \rangle$ much higher than $\mathrm{STARS^{SS}}$.
Due to the hardness of the BS spectra, the relative difference of $\langle x_{\rm HeIII} \rangle$ between simulations $\mathrm{STARS^{BS}}$ and $\mathrm{STARS^{SS}}$ is maintained above $200\%$ at $z<9$. As the effect of BS is dampened by the presence of energetic sources (e.g.  $\mathrm{ALL^{SS}}$ produce a $\langle x_{\rm HeIII} \rangle$ $\sim 30/160$ times higher than that of $\mathrm{STARS^{SS}}$ at $z = 7.5/6.5$), the ratio of $\langle x_{\rm HeIII} \rangle$ between $\mathrm{ALL^{BS}}$ and $\mathrm{ALL^{SS}}$ is smaller, the maximum being $\sim 1.3$ at $z = 8$. 

By assuming that helium is fully ionized below $z=3$, the reionization histories discussed above produce a Thomson scattering optical depth to CMB photons $\tau \approx 0.057$ in simulations with SS, while $\tau \approx 0.062$ in simulations with BS. 
As a comparison, the one measured by the Planck telescope is $ 0.054 \pm 0.007$ \citep{Planck2020}, where the confidence region is 68\%.
This means that despite the earlier reionization reached in the presence of BS, the results of all our simulations are still consistent with CMB observations.

\begin{figure*}
\centering
	\includegraphics[width=0.96\linewidth]{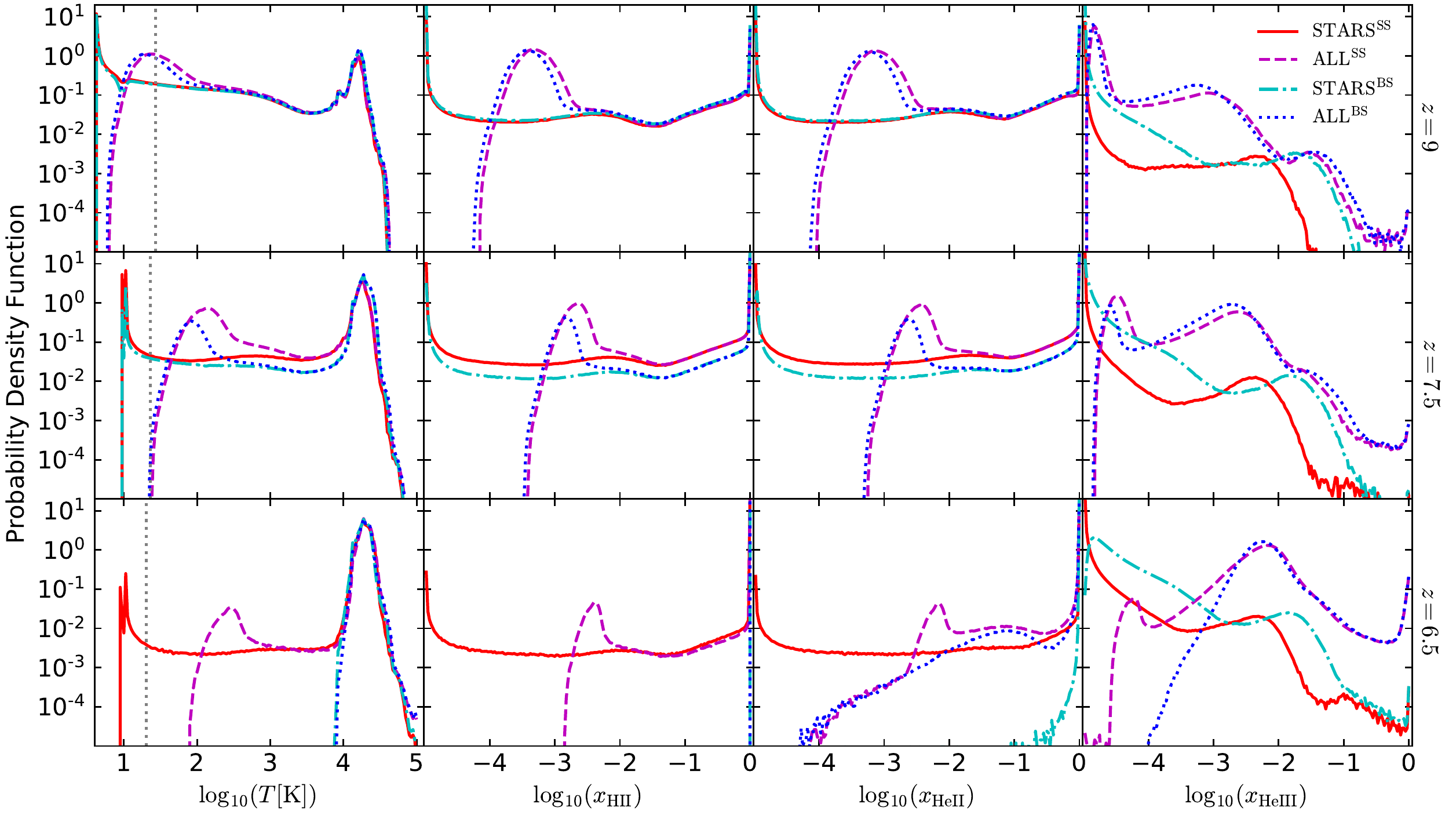}
    \caption{From left to right, probability density functions of gas temperature $T$, HII fraction $x_{\rm HII}$, HeII fraction $x_{\rm HeII}$ and HeIII fraction $x_{\rm HeIII}$ of cells from simulations $\mathrm{STARS^{SS}}$ (red solid lines), $\mathrm{ALL^{SS}}$ (magenta dashed), $\mathrm{STARS^{BS}}$ (cyan dash-dotted) and $\mathrm{ALL^{BS}}$ (blue dotted). 
    From top to bottom the rows refer to $z$ = 9, 7.5 and 6.5. 
    The gray dotted vertical lines in the first column denotes the CMB temperature at the corresponding redshift. 
    }
    \label{fig:frac1}
\end{figure*} 
In Fig.~\ref{fig:frac1} we present the 1-D Probability Density Functions (PDF) of cell properties from our four simulations at $z = 9$, 7.5 and 6.5.
As a reference, in Fig.~\ref{fig:frac1_vi} of appendix \ref{appendix:pro_vi} we show the PDFs at the same HII fraction, i.e. $\langle x_{\rm HII} \rangle = 0.1$, 0.5 and 0.9.
At $z = 9$ and 7.5, the temperature curves show two peaks. The one at $T\sim 10^{4.2}\,\rm K$ is associated with fully ionized cells and is thus similar in all simulations. 
As more ionized cells are produced in the presence of BS, the amplitude of the peak is more pronounced in $\mathrm{STARS^{BS}}$ and $\mathrm{ALL^{BS}}$ than in $\mathrm{STARS^{SS}}$ and $\mathrm{ALL^{SS}}$.
The second peak is associated with neutral (in STARS simulations) or mildly ionized (in ALL simulations) cells. 
As discussed in E20, the heating from energetic sources shifts this peak to temperature values higher than in the STARS simulations, and it keeps increasing with decreasing redshift, e.g. from $T = 20\,\rm K$ at $z = 9$ to $T = 100\,\rm K$ (larger than the CMB temperature $T_{\rm CMB} \approx 23\,\rm K$) at $z = 7.5$.
Due to the more advanced ionization, the amplitude of the peak with BS is slightly lower than that with SS. 
The distributions of the low temperature tail in simulations $\mathrm{ALL^{BS}}$ and $\mathrm{ALL^{SS}}$ are very similar, e.g. at $T<100\,\rm K$ and $z = 7.5$, confirming that BS do not affect the heating of the neutral IGM.
At $z = 6.5$, reionization is complete in the presence of BS and thus the corresponding temperature presents a Gaussian-like PDF. 
Meanwhile, a small fraction of cells in the simulations with SS are still cold.

Since the HII and HeII ionization pattern in all four simulations is similar (see Fig.~\ref{fig:GAL_z_7.5} and Fig.~\ref{fig:ALL_z_7.5}), $x_{\rm HII}$ and $x_{\rm HeII}$ present also a similar PDF at $z=9$ and 7.5, which resembles the behaviour observed in the temperature curves, with a peak corresponding to fully ionized gas ($x_{\rm HII} \approx x_{\rm HeII} \sim 1$), and a second corresponding to neutral (in STARS, $x_{\rm HII} \approx x_{\rm HeII} \sim 10^{-5}$) or partially ionized (in ALL, $x_{\rm HII} \approx x_{\rm HeII} \sim 10^{-3.3}$ at $z = 9$ and $\sim 10^{-2.7}$ at $z = 7.5$) gas.
At $z=6.5$ some differences can be observed. 
In the simulations with BS, the hydrogen reionization is complete, i.e. all cells have $x_{\rm HII} \sim 1$. 
As BS are able to ionize HeII, the cells in simulation $\mathrm{STARS^{BS}}$ have $x_{\rm HeII}$ as low as $\sim 0.1$. 
The efficiency of energetic sources (mostly BHs, see E20) at fully ionizing helium becomes clearly visible, resulting in an even lower abundance of HeII in $\mathrm{ALL^{BS}}$.
The HeII peak at $\lesssim 10^{-2.2}$ in $\mathrm{ALL^{SS}}$ is not present in $\mathrm{ALL^{BS}}$ due to the much reduced partially ionized IGM in the latter case.
The progression of reionization is also observed in the simulations with SS, e.g. a larger number of fully ionized cells, a reduction of highly neutral cells in STARS, and a higher partial ionization of neutral cells in ALL.

The four simulations show more pronounced differences in the $x_{\rm HeIII}$ PDFs. 
In the STARS simulations we observe two peaks. 
The one at $x_{\rm HeIII} \sim 10^{-5}$ is associated with cells where very little HeII is ionized, which is similar with BS and SS. 
The second peak, instead, corresponds to higher values of $x_{\rm HeIII}$ in the presence of BS, e.g. at $z=9$ it is $\sim 10^{-1.7}$ and $\sim 10^{-2.3}$ with BS and SS, respectively. 
The location of the peak is more or less maintained at lower redshift, although the amplitude is increased since more HeII is ionized. 
As mentioned above, the effect of BS is less visible in the presence of more energetic sources. 
Indeed the PDFs for the two ALL simulations are very similar at $z = 9$ and 7.5, with a peak at $x_{\rm HeIII} < 10^{-4}$ (dominated by the almost neutral cells), one at $\sim 10^{-1.5}$ (corresponding to cells close to the sources, where the ionization flux is stronger and HeII more easily ionized), and an intermediate one at $\sim  10^{-3}$ (corresponding to cells ionized but further away from the sources).
At $z = 6.5$, $\mathrm{ALL^{SS}}$ has a PDF amplitude at $x_{\rm HeIII} <  10^{-3}$ higher than $\mathrm{ALL^{BS}}$, since in the former reionization of HeI is not complete yet and partially ionized cells have also a low HeIII ionization, i.e.  $x_{\rm HeIII} \sim  10^{-4.2}$. 

\subsection{21 cm signal}
\label{21cm_sig}
With the information on gas ionization, temperature and density discussed above, we compute the 21 cm differential brightness temperature (DBT) following \citet[][hereafter Ma21]{Ma2021}. 
As in Ma21, we include redshift-space distortion effects with the approach of \cite{Mao2012}, as well as the temperature corrections necessary to overcome the impossibility of resolving the ionization front in such large scale simulations.
Such technique has been introduced and  described in \cite{Ma2020}. 
As discussed in Ma21, the temperature correction not only changes the statistics of the 21 cm signal during the early stages of the EoR (e.g. in our ALL simulations), but also those with weak heating at lower redshift (e.g. our STARS simulations).
Similarly to Ma21, we normalize the 21 cm power spectra (PS) as $\Delta_{\rm 21cm} (k) = k^{3}/2\pi^{2} \times P_{\rm 21cm}(k)$, where $P_{\rm 21cm} (k) = \langle \delta_{\rm D}({\bm k} +{\bm k'})  T_{\rm 21cm}({\bm k}) T_{\rm 21cm}({\bm k'}) \rangle$, $\delta_{\rm D}$ is the Dirac function and $T_{\rm 21cm}({\bm k})$ is the 21 cm DBT in Fourier space. 
We additionally assume that the spin temperature $T_{s}$ is coupled to the gas kinetic temperature $T_{k}$. 
As shown in Ma21, this assumption should not significantly affect the 21 cm results at $z<13$.
We also note that, due to sample variance, the results of the 21 cm PS are not robust at large scales (e.g. $k < 0.3\,\rm Mpc^{-1}$ as estimated in Ma21).

\begin{figure*}
\centering
	\includegraphics[width=0.92\linewidth]{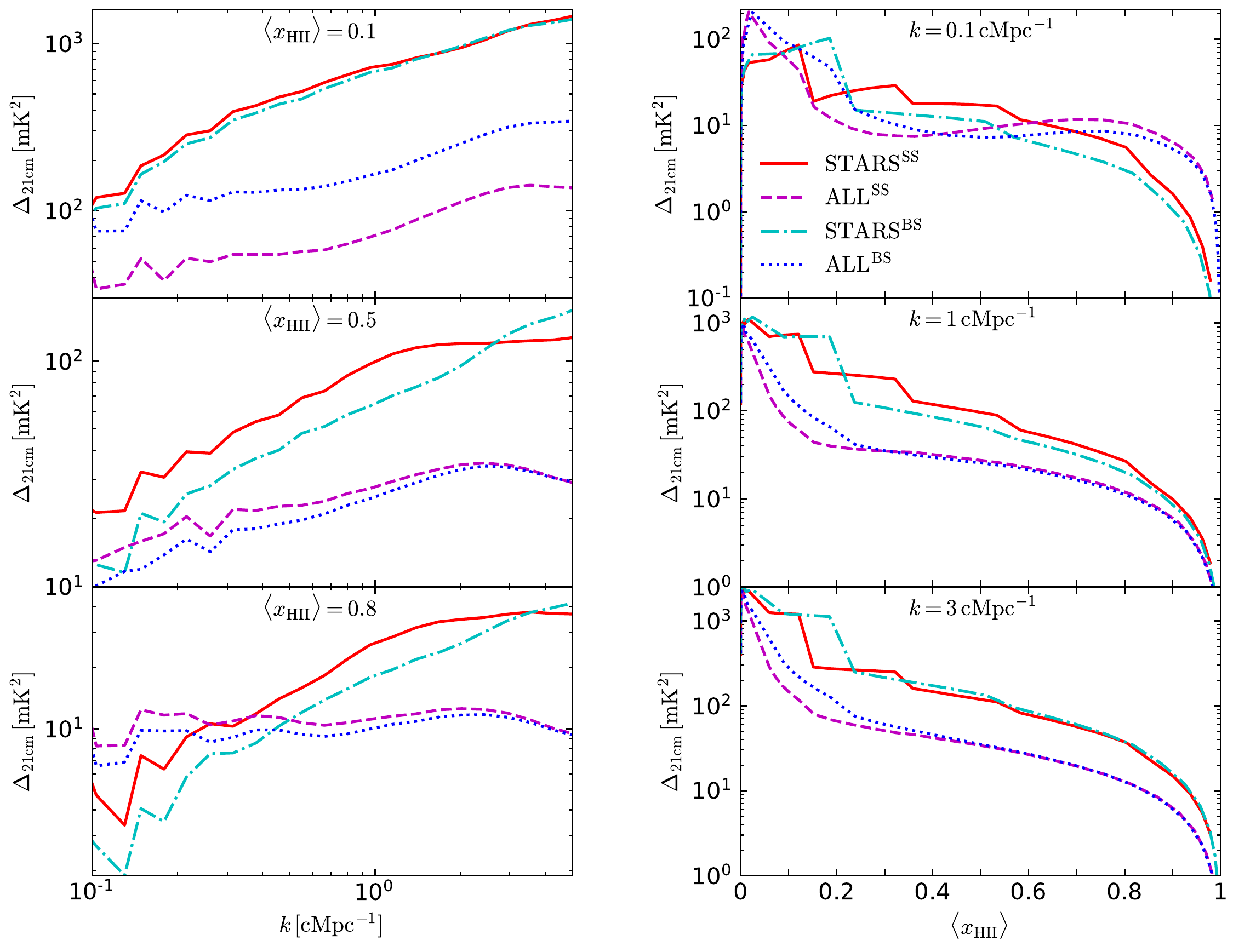}
    \caption{21~cm power spectra $\Delta_{\rm 21cm}$ from simulations $\mathrm{STARS^{SS}}$ (red solid lines), $\mathrm{ALL^{SS}}$ (magenta dashed), $\mathrm{STARS^{BS}}$ (cyan dash-dotted) and $\mathrm{ALL^{BS}}$ (blue dotted). 
    The left panel shows $\Delta_{\rm 21cm}$ at, from top to bottom, $\langle x_{\rm HII} \rangle = 0.1$, 0.5 and 0.8.
    The right panel is the evolution of $\Delta_{\rm 21cm}$ as a function of $\langle x_{\rm HII} \rangle$, at $k = 0.1$, 1 and 3$\,\rm cMpc^{-1}$, from top to bottom.
    }
    \label{fig:ps_21cm_vi}
\end{figure*} 
The left panels of Fig.~\ref{fig:ps_21cm_vi} show  $\Delta_{\rm 21cm}$ at $\langle x_{\rm HII} \rangle = 0.1$ (corresponding to $z \sim 9.2$ and 9.7 for SS and BS simulations, respectively), 0.5 ($z \sim 7.6$ and 8) and 0.8 ($z \sim 7.0$ and 7.5).
Note that the jagged behaviour observed at large scales is due to the poor sampling occurring on scales approaching those of the box size.
During the early stages of the EoR (shown in the figure at $\langle x_{\rm HII} \rangle = 0.1$),
the fluctuations of 21 cm DBT without heating from energetic sources are dominated by the gas density, thus $\mathrm{STARS^{SS}}$ and $\mathrm{STARS^{BS}}$ have a similar $\Delta_{\rm 21cm}$.
The effect of heating is visible in terms of a reduced amplitude of the PS (see Ma21 for a more extended discussion), which is more pronounced on the smaller scales.
Since $\langle x_{\rm HII} \rangle \sim 0.1$ is reached in $\mathrm{ALL^{SS}}$  later than in $\mathrm{ALL^{BS}}$, in the former simulation X-ray heating is active for a longer time, resulting into the amplitude of $\Delta_{\rm 21cm}$ several times smaller in $\mathrm{ALL^{SS}}$ than in $\mathrm{ALL^{BS}}$, although the shape of their spectra looks similar.

As reionization proceeds, e.g. at $\langle x_{\rm HII} \rangle = 0.5$ and 0.8, the fluctuations in the ionization field start to influence the PS in the STARS simulations, and the effect of BS becomes more complex, e.g. $\mathrm{STARS^{SS}}$ presents larger $\Delta_{\rm 21cm}$ at $k$ below a few cMpc$^{-1}$ (depending on $\langle x_{\rm HII} \rangle$). 
A similar behaviour is observed in the presence of energetic sources, i.e.  $\mathrm{ALL^{SS}}$ has a $\Delta_{\rm 21cm}$ larger than $\mathrm{ALL^{BS}}$.
We perform a perturbative analysis (i.e. Taylor expansion) of the 21 cm signal \citep{Barkana2005, Santos2005, Mesinger2019book} by writing $P_{\rm 21cm} = P_{\delta\delta} + P_{\delta_{x}\delta_{x}} + P_{\delta_{T}\delta_{T}} + {c.c}$, where $P_{\delta\delta}$, $P_{\delta_{x}\delta_{x}}$ and $P_{\delta_{T}\delta_{T}}$ are the components of matter density, ionization fraction and spin temperature of the 21 cm power spectrum respectively, while $c.c$ denotes the parts of their cross-correlations.
We find that such behaviour of $\Delta_{\rm 21cm}$ is caused by the anti-correlation of neutral hydrogen fraction with matter density and spin temperature, which happens when reionization is well under way. 
As mentioned earlier, since at the time corresponding to the same ionization fraction heating has been active longer in simulations with SS than in those with BS, this leads to higher fluctuations of the spin temperature for the latter case, but their anti-correlation with the neutral hydrogen fraction has a dominant effect and results in a lower amplitude of $\Delta_{\rm 21cm}$ in simulations with BS than in those with SS. 
We highlight that, although the differences induced by the presence of BS are smaller than those introduced by energetic sources, they are still clearly visible. 

The right panels of Fig.~\ref{fig:ps_21cm_vi} show the evolution of $\Delta_{\rm 21cm}$ with $\langle x_{\rm HII} \rangle$ at $k = 0.1$, 1 and 3$\,\rm cMpc^{-1}$.
At small scales, i.e. $k = 3 \,\rm cMpc^{-1}$, simulations with the same heating source model show a similar $\Delta_{\rm 21cm}$, and the PS of the STARS simulations are $\sim 2$ times higher than those of the ALL simulations. 
Some differences, though, are visible among the ALL simulations at $\langle x_{\rm HII} \rangle \lesssim 0.3$. 
As already mentioned, this is because the earlier ionization by BS results in less X-ray heating in $\mathrm{ALL^{BS}}$ and thus a higher $\Delta_{\rm 21cm}$. 
At larger scales (shown here at $k = 0.1 \,\rm cMpc^{-1}$) the four simulations present more visible differences.
Similar to that at $k = 3 \,\rm cMpc^{-1}$, the $\Delta_{\rm 21cm}$ of $\mathrm{ALL^{BS}}$ is larger than the one from $\mathrm{ALL^{SS}}$ at an early stage of the EoR, i.e. $\langle x_{\rm HII} \rangle < 0.4$.
Differently, at $\langle x_{\rm HII} \rangle > 0.4$, $\mathrm{ALL^{BS}}$ has a PS smaller than that of $\mathrm{ALL^{SS}}$.
As discussed for the left panels of Fig.~\ref{fig:ps_21cm_vi}, this is due to the anti-correlation of spin temperature and neutral hydrogen fraction during the intermediate and later stages of the EoR, which reduces the fluctuations of 21 cm DBT at large scales. 
For the same reason, $\mathrm{STARS^{SS}}$ has a PS larger than $\mathrm{STARS^{BS}}$ at $\langle x_{\rm HII} \rangle > 0.2$.
The behaviour of $\Delta_{\rm 21cm}$ at $k = 1 \,\rm cMpc^{-1}$ is between that at $k = 0.1$ and at $3 \,\rm cMpc^{-1}$, e.g. $\mathrm{STARS^{SS}}$ displays a  slightly higher $\Delta_{\rm 21cm}$ than $\mathrm{STARS^{BS}}$ at $\langle x_{\rm HII} \rangle > 0.2$, but the difference is not as pronounced as that at $k = 0.1 \,\rm cMpc^{-1}$. 
Also here it is clear that the differences caused by the energetic sources are more significant than those due to the presence of BS.

\subsection{Topology of ionized bubbles}
\label{sec:topology}
While the effect of binary stars on the 21 cm signal might be hidden by the impact of energetic sources, in this section we investigate if it remains visible in the topology of ionized regions. 

To do that, we use the recent and versatile pipeline developed by \citealt{Busch2020} (hereafter B20), which employs a binary representation of the ionization fraction fields of hydrogen and helium obtained by imposing an ionization threshold, which we assume to be  $x_{\rm HII}^{\rm th} = 0.99$. 
We have verified that adopting $x_{\rm HII}^{\rm th} = 0.1$ does not affect the following discussion, while even smaller values, e.g. $<0.01$, are expected to result in a different behaviour as the partial ionization from energetic sources would be detected. 
The binary images obtained are thus processed with the morphological opening transform using a series of spherical structure elements of increasing radius. 
The bubbles are then deconstructed into a minimal set of overlapping structures, while the maximally large spherical regions are eventually used to understand the bubble sizes, the arrangements, the connectivity, as well as the density profiles. 
This technique, also called granulometry, was firstly introduced in \cite{Kakiichi2017}, and further developed by B20 to study the topology of ionized bubbles. 
We thus refer the reader to \cite{Kakiichi2017} and B20 for further details.

\begin{figure*}
\centering
	\includegraphics[width=0.96\linewidth]{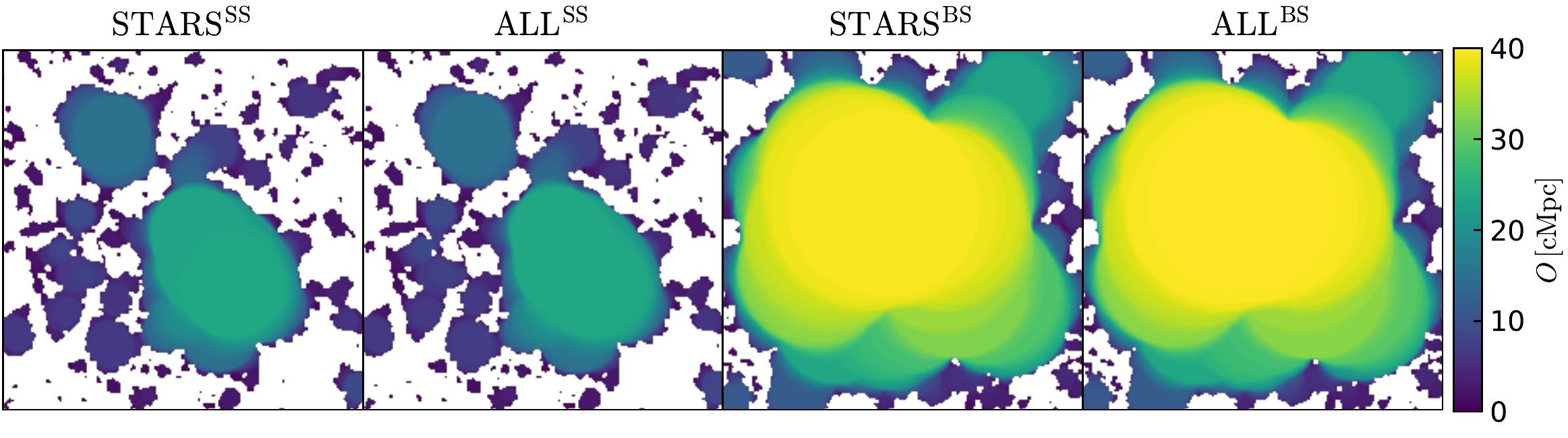}
    \caption{Maps of opening values corresponding to the ionization fields shown in  Fig.~\ref{fig:GAL_z_7.5} and Fig.~\ref{fig:ALL_z_7.5}. 
    From left to right, the results of the simulation $\mathrm{STARS^{SS}}$, $\mathrm{ALL^{SS}}$, $\mathrm{STARS^{BS}}$ and $\mathrm{ALL^{BS}}$. 
    }
    \label{fig:bubble_op_images}
\end{figure*}
As a reference, in Fig.~\ref{fig:bubble_op_images} we show the results of the granulometry applied to our four simulations at $z = 7.5$, i.e. the images correspond to the  $x_{\rm HII}$ ones shown in Fig.~\ref{fig:GAL_z_7.5} and Fig.~\ref{fig:ALL_z_7.5}. 
Note that, as a result of the ionization threshold imposed, the ionized regions resulting from the granulometric analysis look slightly smaller than those in Fig.~\ref{fig:GAL_z_7.5} and Fig.~\ref{fig:ALL_z_7.5}.
At each location, the opening value denotes the radius of the largest spherical ionized region to which the cell belongs. As highly ionized gas is dominated by stellar type sources, the STARS and ALL simulations have an almost identical opening field, while the inclusion of BS increases the maximum opening value from $24\,\rm cMpc$ to $40\,\rm cMpc$. 

\begin{figure}
\centering
	\includegraphics[width=0.92\linewidth]{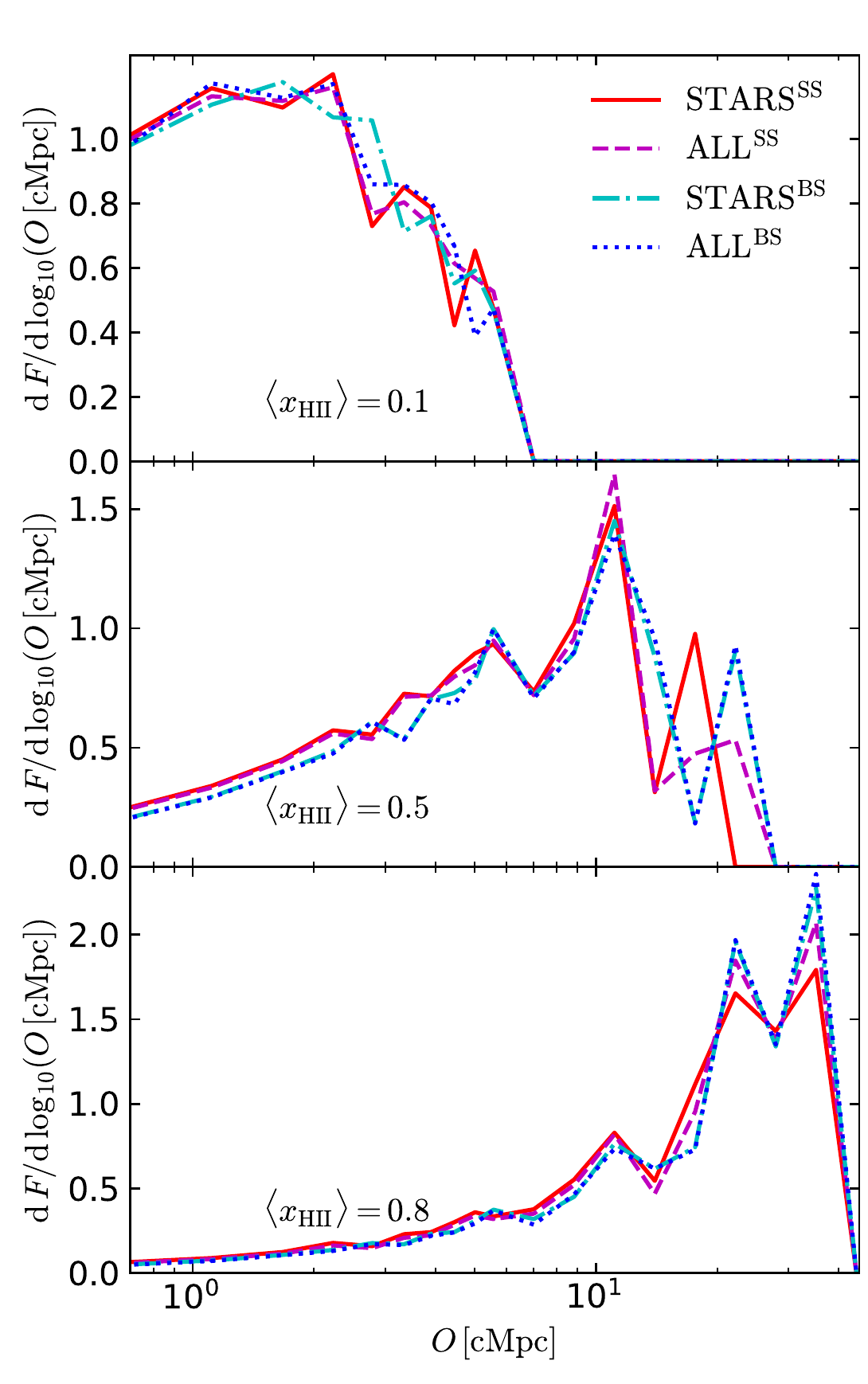}
    \caption{Volume weighted probability distributions (i.e. ${\rm d}\,F/{\rm d\, log_{10}}\,O$ of being in a region with $x_{\rm HII} > 0.99$) as a function of the opening value $O$, from simulation $\mathrm{STARS^{SS}}$ (red solid lines), $\mathrm{ALL^{SS}}$ (magenta dashed), $\mathrm{STARS^{BS}}$ (cyan dash-dotted) and $\mathrm{ALL^{BS}}$ (blue dotted). From top to bottom, the results are at $\langle x_{\rm HII} \rangle = 0.1$, 0.5 and 0.8, respectively.}
    \label{fig:bubble_op}
\end{figure}
In Fig.~\ref{fig:bubble_op}, we show the volume weighted probability distribution, ${\rm d}\,F/{\rm d}\,O$, for an opening value, $O$, to be in a region with radius $R$. Here $F$ is the pattern spectrum of granulometry (i.e. the Eq.~6 in B20), and $O$ is the bin of radius $R$, i.e. $R \leq O \leq R+\Delta R$.
Such distributions indicate which size of the ionized regions dominates in the various simulations, i.e. the position of the distribution peaks along the x-axis denotes the typical sizes of ionized bubbles. 
From Fig.~\ref{fig:bubble_op}, it is clear that the typical size of ionized bubbles increases quickly with increasing $\langle x_{\rm HII} \rangle$, e.g. it is only $\sim 2\, {\rm cMpc}$ at $\langle x_{\rm HII} \rangle = 0.1$, $\sim 11\, {\rm cMpc}$ at $\langle x_{\rm HII} \rangle = 0.5$, while it grows to about $35\, {\rm cMpc}$ at $\langle x_{\rm HII} \rangle = 0.8$ (see also the discussions in B20).
Since full hydrogen ionization is dominated by stellar sources, the simulations with STARS and ALL show almost the same distributions.
Although it is not visible at $\langle x_{\rm HII} \rangle = 0.1$, the simulations with BS have less small size bubbles (e.g. by $\sim 8\%$ at $O<10\, {\rm cMpc}$ and $\langle x_{\rm HII} \rangle = 0.5$, and by $\sim 10\%$ at $O<20\, {\rm cMpc}$ and $\langle x_{\rm HII} \rangle = 0.8$) than those with SS, while they have slightly more large size bubbles. 
This is due to the higher UV emission of BS that ionizes neutral gas at higher $z$, when the number density of stellar sources is smaller than that at lower $z$.

\section{Discussion and Conclusions}
\label{Sec4:con}

We used the high-resolution hydrodynamical cosmological simulation MassiveBlack-II and four 3D multi-frequency Monte Carlo radiative transfer (CRASH) simulations to investigate the effects of binary stars on the physical properties of the IGM, i.e. the ionization state of hydrogen and helium, and the gas temperature. 
In addition to using two simulations from \cite{Eide2018} and \cite{Eide2020}, i.e. one with only single stars ($\rm STARS^{SS}$), and one including also XRBs, ISM and BHs ($\rm ALL^{SS}$), we run two new simulations which include the 
contribution of binary stars to that of single stars ($\rm STARS^{BS}$), and of all combined sources ($\rm ALL^{BS}$).
With these, we also studied the effects of binary stars on the 21 cm signal and the topology of ionized bubbles.  

We confirmed that binary stars speed up the reionization process of both H and He \citep{Rosdahl2018, Doughty2021}, due to their higher UV emission and harder spectra.
Although, differently from SS, BS are able to produce HeIII in appreciable quantities, the ionization of HeII is dominated by energetic sources, especially the black holes.
We also found that simulations with BS result in globally higher IGM temperature \cite[see also ][]{Doughty2021},
while the behaviour within the fully ionized regions is more complex.
Indeed, since the earlier ionization of cells in the presence of BS results in more time for cooling and recombination, many fully ionized cells in the simulations with BS show a temperature lower than that of the same cells in the simulations with SS. This depends, though, also on the relative locations of the cells with respect to the sources, as in their vicinity the ionizing flux is stronger and additionally BS are able to fully ionize helium, resulting in higher temperatures. 
Differently, BS have no obvious effects on the temperature and ionization state of the neutral gas surrounding fully ionized regions, which is instead dominated by the energetic sources.

For the same volume averaged ionization fraction, $\langle x_{\rm HII} \rangle$, the 21~cm power spectra are visibly changed by the presence of BS, although the differences are much smaller than those caused by energetic sources.
The effect of BS is only visible at large scales (e.g. $k = 0.1\,\rm cMpc^{-1}$), while it is almost negligible at small scales (e.g. $k > 1\,\rm cMpc^{-1}$).

From a topological analysis of the ionized bubbles, we find that at the same $\langle x_{\rm HII} \rangle$ simulations with BS have fewer small size bubbles but more large bubbles, although the difference is not extremely significant (e.g. $\sim 8\%$ for radii below $10\, {\rm cMpc}$ and $\langle x_{\rm HII} \rangle = 0.5$), indeed it is smaller than the SKA error-bars estimated in e.g. \cite{Kakiichi2017}, rendering the detection of such differences very challenging.

As in the current simulations the radiative transfer is followed in post-processing, feedback due to ionization and heating from UV and X-ray sources is not accounted for self-consistently. 
While this is expected to have a local effect on e.g. the star formation efficiency in low-mass galaxies, the global effect on the reionization process is somewhat degenerate with the escape fraction of ionizing photons. 
This argument applies also to the specific case of BS, whose presence is expected to reduce the star formation rate in galaxies with stellar mass below $10^8$~M$_\odot$ \cite[e.g.][]{Doughty2021}, but at the same time to increase the escape fraction of ionizing photons \cite[e.g.][]{MaX2016,MaX2020,Secunda2020}. 
A quantitative evaluation of the net effect would require dedicated and specifically tailored simulations, which is beyond the scope of this work.
We also note that the binary interaction algorithm depends on the initial mass function (IMF) of stars, the initial separation of the two stars and their mass ratio. While the IMF is chosen self-consistently with the one adopted in the hydrodynamic simulations, we otherwise used the results from the \cite{Eldridge2012} fiducial model, i.e varying other parameters of the stellar population synthesis code might change our quantitative conclusions, while we expect qualitative considerations to hold. 
Besides, the net effects of binary stars might be degenerate with a higher efficiency of single star formation, although some differences should still be visible in the ionization of Helium, which is sensitive to the harder spectrum of binary stars.

Our main conclusions can be summarized as follows:
\begin{itemize}
    \item Binary stars contribute more (and more energetic) ionizing photons than single stars, thus speeding up the reionization of HI and HeI;
    \item ionization of HeII during the EoR is dominated by energetic sources (in particular black holes, but also shock heated ISM and X-ray binaries), although the harder spectrum of binary stars also contribute abundant photons above $54.4\,\rm eV$;
    \item at any given redshift, ionized gas in the case of binary stars is typically colder than in that of single stars, since it gets ionized earlier and thus has more time to cool and recombine. The globally volume averaged  temperature of the former case is instead higher;
    \item binary stars have a clear effect on the 21 cm power spectra, although the differences compared to those with single stars are smaller than those caused by the more energetic sources;
    \item at the same ionization fraction, models including binary stars result in fewer small size ionized bubbles, but more large ones.
\end{itemize}

\section*{Acknowledgements}
SF is grateful to Martin Glatzle and Enrico Garaldi for insightful discussions and constant support.
We thank an anonymous referee for her/his useful comments, and Tiziana di Matteo and Yue Feng for providing outputs of the MB-II simulation.
This work is supported by the National SKA Program
of China (grant No. 2020SKA0110402), National Natural Science Foundation of China (Grant No. 11903010), innovation and entrepreneurial project of Guizhou province for high-level overseas talents (Grant No. (2019)02), Science and Technology Fund of Guizhou Province (Grant No. [2020]1Y020), and GZNU 2019 Special projects of training new academics and innovation exploration.
The tools for bibliographic research are offered by the NASA Astrophysics Data Systems and by the JSTOR archive.

\section*{Data availability}
The data of simulations and analysis scripts underlying this article will be shared on reasonable request to the corresponding author.



\bibliographystyle{mnras}
\bibliography{ref} 

\begin{thebibliography}{}
\makeatletter
\relax
\def\mn@urlcharsother{\let\do\@makeother \do\$\do\&\do\#\do\^\do\_\do\%\do\~}
\def\mn@doi{\begingroup\mn@urlcharsother \@ifnextchar [ {\mn@doi@}
  {\mn@doi@[]}}
\def\mn@doi@[#1]#2{\def\@tempa{#1}\ifx\@tempa\@empty \href
  {http://dx.doi.org/#2} {doi:#2}\else \href {http://dx.doi.org/#2} {#1}\fi
  \endgroup}
\def\mn@eprint#1#2{\mn@eprint@#1:#2::\@nil}
\def\mn@eprint@arXiv#1{\href {http://arxiv.org/abs/#1} {{\tt arXiv:#1}}}
\def\mn@eprint@dblp#1{\href {http://dblp.uni-trier.de/rec/bibtex/#1.xml}
  {dblp:#1}}
\def\mn@eprint@#1:#2:#3:#4\@nil{\def\@tempa {#1}\def\@tempb {#2}\def\@tempc
  {#3}\ifx \@tempc \@empty \let \@tempc \@tempb \let \@tempb \@tempa \fi \ifx
  \@tempb \@empty \def\@tempb {arXiv}\fi \@ifundefined
  {mn@eprint@\@tempb}{\@tempb:\@tempc}{\expandafter \expandafter \csname
  mn@eprint@\@tempb\endcsname \expandafter{\@tempc}}}

\bibitem[\protect\citeauthoryear{{Barkana} \& {Loeb}}{{Barkana} \&
  {Loeb}}{2005}]{Barkana2005}
{Barkana} R.,  {Loeb} A.,  2005, \mn@doi [\apjl] {10.1086/430599}, \href
  {https://ui.adsabs.harvard.edu/abs/2005ApJ...624L..65B} {624, L65}

\bibitem[\protect\citeauthoryear{{Berzin}, {Secunda}, {Cen}, {Menegas}  \&
  {G{\"o}tberg}}{{Berzin} et~al.}{2021}]{Berzin2021}
{Berzin} E.,  {Secunda} A.,  {Cen} R.,  {Menegas} A.,   {G{\"o}tberg} Y.,
  2021, \mn@doi [\apj] {10.3847/1538-4357/ac0af6}, \href
  {https://ui.adsabs.harvard.edu/abs/2021ApJ...918....5B} {918, 5}

\bibitem[\protect\citeauthoryear{{Bouwens} et~al.,}{{Bouwens}
  et~al.}{2015a}]{Bouwens2015b}
{Bouwens} R.~J.,  et~al., 2015a, \mn@doi [\apj] {10.1088/0004-637X/803/1/34},
  \href {https://ui.adsabs.harvard.edu/abs/2015ApJ...803...34B} {803, 34}

\bibitem[\protect\citeauthoryear{{Bouwens}, {Illingworth}, {Oesch}, {Caruana},
  {Holwerda}, {Smit}  \& {Wilkins}}{{Bouwens} et~al.}{2015b}]{Bouwens2015}
{Bouwens} R.~J.,  {Illingworth} G.~D.,  {Oesch} P.~A.,  {Caruana} J.,
  {Holwerda} B.,  {Smit} R.,   {Wilkins} S.,  2015b, \mn@doi [\apj]
  {10.1088/0004-637X/811/2/140}, \href
  {https://ui.adsabs.harvard.edu/abs/2015ApJ...811..140B} {811, 140}

\bibitem[\protect\citeauthoryear{{Busch}, {Eide}, {Ciardi}  \&
  {Kakiichi}}{{Busch} et~al.}{2020}]{Busch2020}
{Busch} P.,  {Eide} M.~B.,  {Ciardi} B.,   {Kakiichi} K.,  2020, \mn@doi
  [\mnras] {10.1093/mnras/staa2599}, \href
  {https://ui.adsabs.harvard.edu/abs/2020MNRAS.498.4533B} {498, 4533}

\bibitem[\protect\citeauthoryear{{Ciardi}, {Ferrara}, {Marri}  \&
  {Raimondo}}{{Ciardi} et~al.}{2001}]{Ciardi2001}
{Ciardi} B.,  {Ferrara} A.,  {Marri} S.,   {Raimondo} G.,  2001, \mn@doi
  [\mnras] {10.1046/j.1365-8711.2001.04316.x}, \href
  {https://ui.adsabs.harvard.edu/abs/2001MNRAS.324..381C} {324, 381}

\bibitem[\protect\citeauthoryear{{Croft}, {Di Matteo}, {Springel}  \&
  {Hernquist}}{{Croft} et~al.}{2009}]{Croft2009}
{Croft} R. A.~C.,  {Di Matteo} T.,  {Springel} V.,   {Hernquist} L.,  2009,
  \mn@doi [\mnras] {10.1111/j.1365-2966.2009.15446.x}, \href
  {https://ui.adsabs.harvard.edu/abs/2009MNRAS.400...43C} {400, 43}

\bibitem[\protect\citeauthoryear{{Dayal} \& {Ferrara}}{{Dayal} \&
  {Ferrara}}{2018}]{Dayal2018}
{Dayal} P.,  {Ferrara} A.,  2018, \mn@doi [\physrep]
  {10.1016/j.physrep.2018.10.002}, \href
  {https://ui.adsabs.harvard.edu/abs/2018PhR...780....1D} {780, 1}

\bibitem[\protect\citeauthoryear{{Degraf}, {Di Matteo}  \& {Springel}}{{Degraf}
  et~al.}{2010}]{Degraf2010}
{Degraf} C.,  {Di Matteo} T.,   {Springel} V.,  2010, \mn@doi [\mnras]
  {10.1111/j.1365-2966.2009.16018.x}, \href
  {https://ui.adsabs.harvard.edu/abs/2010MNRAS.402.1927D} {402, 1927}

\bibitem[\protect\citeauthoryear{{Di Matteo}, {Colberg}, {Springel},
  {Hernquist}  \& {Sijacki}}{{Di Matteo} et~al.}{2008}]{Di2008}
{Di Matteo} T.,  {Colberg} J.,  {Springel} V.,  {Hernquist} L.,   {Sijacki} D.,
   2008, \mn@doi [\apj] {10.1086/524921}, \href
  {https://ui.adsabs.harvard.edu/abs/2008ApJ...676...33D} {676, 33}

\bibitem[\protect\citeauthoryear{{Di Matteo}, {Khandai}, {DeGraf}, {Feng},
  {Croft}, {Lopez}  \& {Springel}}{{Di Matteo} et~al.}{2012}]{Di2012}
{Di Matteo} T.,  {Khandai} N.,  {DeGraf} C.,  {Feng} Y.,  {Croft} R.~A.~C.,
  {Lopez} J.,   {Springel} V.,  2012, \mn@doi [\apjl]
  {10.1088/2041-8205/745/2/L29}, \href
  {https://ui.adsabs.harvard.edu/abs/2012ApJ...745L..29D} {745, L29}

\bibitem[\protect\citeauthoryear{{Doughty} \& {Finlator}}{{Doughty} \&
  {Finlator}}{2021}]{Doughty2021}
{Doughty} C.,  {Finlator} K.,  2021, \mn@doi [\mnras] {10.1093/mnras/stab1448},
  \href {https://ui.adsabs.harvard.edu/abs/2021MNRAS.505.2207D} {505, 2207}

\bibitem[\protect\citeauthoryear{{Eide}, {Graziani}, {Ciardi}, {Feng},
  {Kakiichi}  \& {Di Matteo}}{{Eide} et~al.}{2018}]{Eide2018}
{Eide} M.~B.,  {Graziani} L.,  {Ciardi} B.,  {Feng} Y.,  {Kakiichi} K.,   {Di
  Matteo} T.,  2018, \mn@doi [\mnras] {10.1093/mnras/sty272}, \href
  {https://ui.adsabs.harvard.edu/abs/2018MNRAS.476.1174E} {476, 1174}

\bibitem[\protect\citeauthoryear{{Eide}, {Ciardi}, {Graziani}, {Busch}, {Feng}
  \& {Di Matteo}}{{Eide} et~al.}{2020}]{Eide2020}
{Eide} M.~B.,  {Ciardi} B.,  {Graziani} L.,  {Busch} P.,  {Feng} Y.,   {Di
  Matteo} T.,  2020, \mn@doi [\mnras] {10.1093/mnras/staa2774}, \href
  {https://ui.adsabs.harvard.edu/abs/2020MNRAS.498.6083E} {498, 6083}

\bibitem[\protect\citeauthoryear{{Eldridge} \& {Stanway}}{{Eldridge} \&
  {Stanway}}{2012}]{Eldridge2012}
{Eldridge} J.~J.,  {Stanway} E.~R.,  2012, \mn@doi [\mnras]
  {10.1111/j.1365-2966.2011.19713.x}, \href
  {https://ui.adsabs.harvard.edu/abs/2012MNRAS.419..479E} {419, 479}

\bibitem[\protect\citeauthoryear{{Fan}, {Carilli}  \& {Keating}}{{Fan}
  et~al.}{2006}]{Fan2006}
{Fan} X.,  {Carilli} C.~L.,   {Keating} B.,  2006, \mn@doi [\araa]
  {10.1146/annurev.astro.44.051905.092514}, \href
  {http://adsabs.harvard.edu/abs/2006ARA%26A..44..415F} {44, 415}

\bibitem[\protect\citeauthoryear{{Field}}{{Field}}{1959}]{Field1959}
{Field} G.~B.,  1959, \mn@doi [\apj] {10.1086/146653}, \href
  {https://ui.adsabs.harvard.edu/abs/1959ApJ...129..536F} {129, 536}

\bibitem[\protect\citeauthoryear{{Furlanetto}, {Oh}  \& {Briggs}}{{Furlanetto}
  et~al.}{2006}]{Furlanetto2006}
{Furlanetto} S.~R.,  {Oh} S.~P.,   {Briggs} F.~H.,  2006, \mn@doi [\physrep]
  {10.1016/j.physrep.2006.08.002}, \href
  {https://ui.adsabs.harvard.edu/abs/2006PhR...433..181F} {433, 181}

\bibitem[\protect\citeauthoryear{{Glatzle}, {Ciardi}  \& {Graziani}}{{Glatzle}
  et~al.}{2019}]{Glatzle2019}
{Glatzle} M.,  {Ciardi} B.,   {Graziani} L.,  2019, \mn@doi [\mnras]
  {10.1093/mnras/sty2514}, \href
  {https://ui.adsabs.harvard.edu/abs/2019MNRAS.482..321G} {482, 321}

\bibitem[\protect\citeauthoryear{{G{\"o}tberg}, {de Mink}, {Groh}, {Leitherer}
  \& {Norman}}{{G{\"o}tberg} et~al.}{2019}]{Gotberg2019}
{G{\"o}tberg} Y.,  {de Mink} S.~E.,  {Groh} J.~H.,  {Leitherer} C.,   {Norman}
  C.,  2019, \mn@doi [\aap] {10.1051/0004-6361/201834525}, \href
  {https://ui.adsabs.harvard.edu/abs/2019A&A...629A.134G} {629, A134}

\bibitem[\protect\citeauthoryear{{G{\"o}tberg}, {de Mink}, {McQuinn},
  {Zapartas}, {Groh}  \& {Norman}}{{G{\"o}tberg} et~al.}{2020}]{Gotberg2020}
{G{\"o}tberg} Y.,  {de Mink} S.~E.,  {McQuinn} M.,  {Zapartas} E.,  {Groh}
  J.~H.,   {Norman} C.,  2020, \mn@doi [\aap] {10.1051/0004-6361/201936669},
  \href {https://ui.adsabs.harvard.edu/abs/2020A&A...634A.134G} {634, A134}

\bibitem[\protect\citeauthoryear{{Graziani}, {Maselli}  \& {Ciardi}}{{Graziani}
  et~al.}{2013}]{Graziani2013}
{Graziani} L.,  {Maselli} A.,   {Ciardi} B.,  2013, \mn@doi [\mnras]
  {10.1093/mnras/stt206}, \href
  {https://ui.adsabs.harvard.edu/abs/2013MNRAS.431..722G} {431, 722}

\bibitem[\protect\citeauthoryear{{Graziani}, {Ciardi}  \& {Glatzle}}{{Graziani}
  et~al.}{2018}]{Graziani2018}
{Graziani} L.,  {Ciardi} B.,   {Glatzle} M.,  2018, \mn@doi [\mnras]
  {10.1093/mnras/sty1367}, \href
  {https://ui.adsabs.harvard.edu/abs/2018MNRAS.479.4320G} {479, 4320}

\bibitem[\protect\citeauthoryear{{Han}, {Ge}, {Chen}  \& {Chen}}{{Han}
  et~al.}{2020}]{Han2020}
{Han} Z.-W.,  {Ge} H.-W.,  {Chen} X.-F.,   {Chen} H.-L.,  2020, \mn@doi [\raa]
  {10.1088/1674-4527/20/10/161}, \href
  {https://ui.adsabs.harvard.edu/abs/2020RAA....20..161H} {20, 161}

\bibitem[\protect\citeauthoryear{{Hariharan}, {Graziani}, {Ciardi}, {Miniati}
  \& {Bungartz}}{{Hariharan} et~al.}{2017}]{Hariharan2017}
{Hariharan} N.,  {Graziani} L.,  {Ciardi} B.,  {Miniati} F.,   {Bungartz}
  H.~J.,  2017, \mn@doi [\mnras] {10.1093/mnras/stx162}, \href
  {https://ui.adsabs.harvard.edu/abs/2017MNRAS.467.2458H} {467, 2458}

\bibitem[\protect\citeauthoryear{{Kakiichi}, {Graziani}, {Ciardi}, {Meiksin},
  {Compostella}, {Eide}  \& {Zaroubi}}{{Kakiichi} et~al.}{2017}]{Kakiichi2017}
{Kakiichi} K.,  {Graziani} L.,  {Ciardi} B.,  {Meiksin} A.,  {Compostella} M.,
  {Eide} M.~B.,   {Zaroubi} S.,  2017, \mn@doi [\mnras] {10.1093/mnras/stx603},
  \href {https://ui.adsabs.harvard.edu/abs/2017MNRAS.468.3718K} {468, 3718}

\bibitem[\protect\citeauthoryear{{Katz}, {Kimm}, {Haehnelt}, {Sijacki},
  {Rosdahl}  \& {Blaizot}}{{Katz} et~al.}{2018}]{Katz2018}
{Katz} H.,  {Kimm} T.,  {Haehnelt} M.,  {Sijacki} D.,  {Rosdahl} J.,
  {Blaizot} J.,  2018, \mn@doi [\mnras] {10.1093/mnras/sty1225}, \href
  {https://ui.adsabs.harvard.edu/abs/2018MNRAS.478.4986K} {478, 4986}

\bibitem[\protect\citeauthoryear{{Keating}, {Puchwein}  \&
  {Haehnelt}}{{Keating} et~al.}{2018}]{Keating2018}
{Keating} L.~C.,  {Puchwein} E.,   {Haehnelt} M.~G.,  2018, \mn@doi [\mnras]
  {10.1093/mnras/sty968}, \href
  {https://ui.adsabs.harvard.edu/abs/2018MNRAS.477.5501K} {477, 5501}

\bibitem[\protect\citeauthoryear{{Khandai}, {Di Matteo}, {Croft}, {Wilkins},
  {Feng}, {Tucker}, {DeGraf}  \& {Liu}}{{Khandai} et~al.}{2015}]{Khandai2015}
{Khandai} N.,  {Di Matteo} T.,  {Croft} R.,  {Wilkins} S.,  {Feng} Y.,
  {Tucker} E.,  {DeGraf} C.,   {Liu} M.-S.,  2015, \mn@doi [\mnras]
  {10.1093/mnras/stv627}, \href
  {https://ui.adsabs.harvard.edu/abs/2015MNRAS.450.1349K} {450, 1349}

\bibitem[\protect\citeauthoryear{{Komatsu} et~al.,}{{Komatsu}
  et~al.}{2011}]{Komatsu2011}
{Komatsu} E.,  et~al., 2011, \mn@doi [\apjs] {10.1088/0067-0049/192/2/18},
  \href {https://ui.adsabs.harvard.edu/abs/2011ApJS..192...18K} {192, 18}

\bibitem[\protect\citeauthoryear{{Krawczyk}, {Richards}, {Mehta}, {Vogeley},
  {Gallagher}, {Leighly}, {Ross}  \& {Schneider}}{{Krawczyk}
  et~al.}{2013}]{Krawczyk2013}
{Krawczyk} C.~M.,  {Richards} G.~T.,  {Mehta} S.~S.,  {Vogeley} M.~S.,
  {Gallagher} S.~C.,  {Leighly} K.~M.,  {Ross} N.~P.,   {Schneider} D.~P.,
  2013, \mn@doi [\apjs] {10.1088/0067-0049/206/1/4}, \href
  {http://adsabs.harvard.edu/abs/2013ApJS..206....4K} {206, 4}

\bibitem[\protect\citeauthoryear{{Loeb} \& {Barkana}}{{Loeb} \&
  {Barkana}}{2001}]{Loeb2001}
{Loeb} A.,  {Barkana} R.,  2001, \mn@doi [\araa]
  {10.1146/annurev.astro.39.1.19}, \href
  {https://ui.adsabs.harvard.edu/abs/2001ARA&A..39...19L} {39, 19}

\bibitem[\protect\citeauthoryear{{Ma}, {Hopkins}, {Kasen}, {Quataert},
  {Faucher-Gigu{\`e}re}, {Kere{\v{s}}}, {Murray}  \& {Strom}}{{Ma}
  et~al.}{2016}]{MaX2016}
{Ma} X.,  {Hopkins} P.~F.,  {Kasen} D.,  {Quataert} E.,  {Faucher-Gigu{\`e}re}
  C.-A.,  {Kere{\v{s}}} D.,  {Murray} N.,   {Strom} A.,  2016, \mn@doi [\mnras]
  {10.1093/mnras/stw941}, \href
  {https://ui.adsabs.harvard.edu/abs/2016MNRAS.459.3614M} {459, 3614}

\bibitem[\protect\citeauthoryear{{Ma}, {Ciardi}, {Eide}  \& {Helgason}}{{Ma}
  et~al.}{2018}]{Ma2018}
{Ma} Q.,  {Ciardi} B.,  {Eide} M.~B.,   {Helgason} K.,  2018, \mn@doi [\mnras]
  {10.1093/mnras/sty1806}, \href
  {https://ui.adsabs.harvard.edu/abs/2018MNRAS.480...26M} {480, 26}

\bibitem[\protect\citeauthoryear{{Ma}, {Quataert}, {Wetzel}, {Hopkins},
  {Faucher-Gigu{\`e}re}  \& {Kere{\v{s}}}}{{Ma} et~al.}{2020a}]{MaX2020}
{Ma} X.,  {Quataert} E.,  {Wetzel} A.,  {Hopkins} P.~F.,  {Faucher-Gigu{\`e}re}
  C.-A.,   {Kere{\v{s}}} D.,  2020a, \mn@doi [\mnras] {10.1093/mnras/staa2404},
  \href {https://ui.adsabs.harvard.edu/abs/2020MNRAS.498.2001M} {498, 2001}

\bibitem[\protect\citeauthoryear{{Ma}, {Ciardi}, {Kakiichi}, {Zaroubi}, {Zhi}
  \& {Busch}}{{Ma} et~al.}{2020b}]{Ma2020}
{Ma} Q.-B.,  {Ciardi} B.,  {Kakiichi} K.,  {Zaroubi} S.,  {Zhi} Q.-J.,
  {Busch} P.,  2020b, \mn@doi [\apj] {10.3847/1538-4357/ab5b95}, \href
  {https://ui.adsabs.harvard.edu/abs/2020ApJ...888..112M} {888, 112}

\bibitem[\protect\citeauthoryear{{Ma}, {Ciardi}, {Eide}, {Busch}, {Mao}  \&
  {Zhi}}{{Ma} et~al.}{2021}]{Ma2021}
{Ma} Q.-B.,  {Ciardi} B.,  {Eide} M.~B.,  {Busch} P.,  {Mao} Y.,   {Zhi} Q.-J.,
   2021, \mn@doi [\apj] {10.3847/1538-4357/abefd5}, \href
  {https://ui.adsabs.harvard.edu/abs/2021ApJ...912..143M} {912, 143}

\bibitem[\protect\citeauthoryear{{Madau} \& {Fragos}}{{Madau} \&
  {Fragos}}{2017}]{Madau2017}
{Madau} P.,  {Fragos} T.,  2017, \mn@doi [\apj] {10.3847/1538-4357/aa6af9},
  \href {https://ui.adsabs.harvard.edu/abs/2017ApJ...840...39M} {840, 39}

\bibitem[\protect\citeauthoryear{{Madau}, {Meiksin}  \& {Rees}}{{Madau}
  et~al.}{1997}]{Madau1997}
{Madau} P.,  {Meiksin} A.,   {Rees} M.~J.,  1997, \mn@doi [\apj]
  {10.1086/303549}, \href
  {https://ui.adsabs.harvard.edu/abs/1997ApJ...475..429M} {475, 429}

\bibitem[\protect\citeauthoryear{{Mao}, {Shapiro}, {Mellema}, {Iliev}, {Koda}
  \& {Ahn}}{{Mao} et~al.}{2012}]{Mao2012}
{Mao} Y.,  {Shapiro} P.~R.,  {Mellema} G.,  {Iliev} I.~T.,  {Koda} J.,   {Ahn}
  K.,  2012, \mn@doi [\mnras] {10.1111/j.1365-2966.2012.20471.x}, \href
  {https://ui.adsabs.harvard.edu/abs/2012MNRAS.422..926M} {422, 926}

\bibitem[\protect\citeauthoryear{{Maselli}, {Ferrara}  \& {Ciardi}}{{Maselli}
  et~al.}{2003}]{Maselli2003}
{Maselli} A.,  {Ferrara} A.,   {Ciardi} B.,  2003, \mn@doi [\mnras]
  {10.1046/j.1365-8711.2003.06979.x}, \href
  {https://ui.adsabs.harvard.edu/abs/2003MNRAS.345..379M} {345, 379}

\bibitem[\protect\citeauthoryear{{Mesinger}}{{Mesinger}}{2019}]{Mesinger2019book}
{Mesinger} A.,  2019, {The Cosmic 21-cm Revolution; Charting the first billion
  years of our universe}, \mn@doi{10.1088/2514-3433/ab4a73.
}

\bibitem[\protect\citeauthoryear{{Mesinger}, {Ferrara}  \&
  {Spiegel}}{{Mesinger} et~al.}{2013}]{Mesinger2013}
{Mesinger} A.,  {Ferrara} A.,   {Spiegel} D.~S.,  2013, \mn@doi [\mnras]
  {10.1093/mnras/stt198}, \href
  {https://ui.adsabs.harvard.edu/abs/2013MNRAS.431..621M} {431, 621}

\bibitem[\protect\citeauthoryear{{Mineo}, {Gilfanov}  \& {Sunyaev}}{{Mineo}
  et~al.}{2012}]{Mineo2012}
{Mineo} S.,  {Gilfanov} M.,   {Sunyaev} R.,  2012, \mn@doi [\mnras]
  {10.1111/j.1365-2966.2012.21831.x}, \href
  {https://ui.adsabs.harvard.edu/abs/2012MNRAS.426.1870M} {426, 1870}

\bibitem[\protect\citeauthoryear{{Morales} \& {Wyithe}}{{Morales} \&
  {Wyithe}}{2010}]{Morales2010}
{Morales} M.~F.,  {Wyithe} J. S.~B.,  2010, \mn@doi [\araa]
  {10.1146/annurev-astro-081309-130936}, \href
  {https://ui.adsabs.harvard.edu/abs/2010ARA&A..48..127M} {48, 127}

\bibitem[\protect\citeauthoryear{{Moriwaki}, {Yoshida}, {Eide}  \&
  {Ciardi}}{{Moriwaki} et~al.}{2019}]{Moriwaki2019}
{Moriwaki} K.,  {Yoshida} N.,  {Eide} M.~B.,   {Ciardi} B.,  2019, \mn@doi
  [\mnras] {10.1093/mnras/stz2308}, \href
  {https://ui.adsabs.harvard.edu/abs/2019MNRAS.489.2471M} {489, 2471}

\bibitem[\protect\citeauthoryear{{Paardekooper}, {Khochfar}  \& {Dalla
  Vecchia}}{{Paardekooper} et~al.}{2015}]{Paardekooper2015}
{Paardekooper} J.-P.,  {Khochfar} S.,   {Dalla Vecchia} C.,  2015, \mn@doi
  [\mnras] {10.1093/mnras/stv1114}, \href
  {https://ui.adsabs.harvard.edu/abs/2015MNRAS.451.2544P} {451, 2544}

\bibitem[\protect\citeauthoryear{{Planck Collaboration} et~al.,}{{Planck
  Collaboration} et~al.}{2020}]{Planck2020}
{Planck Collaboration} et~al., 2020, \mn@doi [\aap]
  {10.1051/0004-6361/201833910}, \href
  {https://ui.adsabs.harvard.edu/abs/2020A&A...641A...6P} {641, A6}

\bibitem[\protect\citeauthoryear{{Price}, {Trac}  \& {Cen}}{{Price}
  et~al.}{2016}]{Price2016}
{Price} L.~C.,  {Trac} H.,   {Cen} R.,  2016, arXiv e-prints, \href
  {https://ui.adsabs.harvard.edu/abs/2016arXiv160503970P} {p. arXiv:1605.03970}

\bibitem[\protect\citeauthoryear{{Roberts-Borsani}, {Treu}, {Mason}, {Schmidt},
  {Jones}  \& {Fontana}}{{Roberts-Borsani} et~al.}{2021}]{Roberts2021}
{Roberts-Borsani} G.,  {Treu} T.,  {Mason} C.,  {Schmidt} K.~B.,  {Jones} T.,
  {Fontana} A.,  2021, \mn@doi [\apj] {10.3847/1538-4357/abe45b}, \href
  {https://ui.adsabs.harvard.edu/abs/2021ApJ...910...86R} {910, 86}

\bibitem[\protect\citeauthoryear{{Rosdahl} et~al.,}{{Rosdahl}
  et~al.}{2018}]{Rosdahl2018}
{Rosdahl} J.,  et~al., 2018, \mn@doi [\mnras] {10.1093/mnras/sty1655}, \href
  {https://ui.adsabs.harvard.edu/abs/2018MNRAS.479..994R} {479, 994}

\bibitem[\protect\citeauthoryear{{Ross}, {Dixon}, {Ghara}, {Iliev}  \&
  {Mellema}}{{Ross} et~al.}{2019}]{Ross2019}
{Ross} H.~E.,  {Dixon} K.~L.,  {Ghara} R.,  {Iliev} I.~T.,   {Mellema} G.,
  2019, \mn@doi [\mnras] {10.1093/mnras/stz1220}, \href
  {https://ui.adsabs.harvard.edu/abs/2019MNRAS.487.1101R} {487, 1101}

\bibitem[\protect\citeauthoryear{{Santos}, {Cooray}  \& {Knox}}{{Santos}
  et~al.}{2005}]{Santos2005}
{Santos} M.~G.,  {Cooray} A.,   {Knox} L.,  2005, \mn@doi [\apj]
  {10.1086/429857}, \href
  {https://ui.adsabs.harvard.edu/abs/2005ApJ...625..575S} {625, 575}

\bibitem[\protect\citeauthoryear{{Secunda}, {Cen}, {Kimm}, {G{\"o}tberg}  \&
  {de Mink}}{{Secunda} et~al.}{2020}]{Secunda2020}
{Secunda} A.,  {Cen} R.,  {Kimm} T.,  {G{\"o}tberg} Y.,   {de Mink} S.~E.,
  2020, \mn@doi [\apj] {10.3847/1538-4357/abaefa}, \href
  {https://ui.adsabs.harvard.edu/abs/2020ApJ...901...72S} {901, 72}

\bibitem[\protect\citeauthoryear{{Springel}, {Di Matteo}  \&
  {Hernquist}}{{Springel} et~al.}{2005}]{Springel2005}
{Springel} V.,  {Di Matteo} T.,   {Hernquist} L.,  2005, \mn@doi [\mnras]
  {10.1111/j.1365-2966.2005.09238.x}, \href
  {https://ui.adsabs.harvard.edu/abs/2005MNRAS.361..776S} {361, 776}

\bibitem[\protect\citeauthoryear{{Stanway} \& {Eldridge}}{{Stanway} \&
  {Eldridge}}{2018}]{Stanway2018}
{Stanway} E.~R.,  {Eldridge} J.~J.,  2018, \mn@doi [\mnras]
  {10.1093/mnras/sty1353}, \href
  {https://ui.adsabs.harvard.edu/abs/2018MNRAS.479...75S} {479, 75}

\bibitem[\protect\citeauthoryear{{Stanway}, {Eldridge}  \& {Becker}}{{Stanway}
  et~al.}{2016}]{Stanway2016}
{Stanway} E.~R.,  {Eldridge} J.~J.,   {Becker} G.~D.,  2016, \mn@doi [\mnras]
  {10.1093/mnras/stv2661}, \href
  {https://ui.adsabs.harvard.edu/abs/2016MNRAS.456..485S} {456, 485}

\bibitem[\protect\citeauthoryear{{Stark}}{{Stark}}{2016}]{Stark2016}
{Stark} D.~P.,  2016, \mn@doi [\araa] {10.1146/annurev-astro-081915-023417},
  \href {https://ui.adsabs.harvard.edu/abs/2016ARA&A..54..761S} {54, 761}

\bibitem[\protect\citeauthoryear{{Trebitsch}, {Blaizot}, {Rosdahl}, {Devriendt}
   \& {Slyz}}{{Trebitsch} et~al.}{2017}]{Trebitsch2017}
{Trebitsch} M.,  {Blaizot} J.,  {Rosdahl} J.,  {Devriendt} J.,   {Slyz} A.,
  2017, \mn@doi [\mnras] {10.1093/mnras/stx1060}, \href
  {https://ui.adsabs.harvard.edu/abs/2017MNRAS.470..224T} {470, 224}

\bibitem[\protect\citeauthoryear{{Wang} et~al.,}{{Wang}
  et~al.}{2021}]{Wang2021}
{Wang} F.,  et~al., 2021, \mn@doi [\apjl] {10.3847/2041-8213/abd8c6}, \href
  {https://ui.adsabs.harvard.edu/abs/2021ApJ...907L...1W} {907, L1}

\bibitem[\protect\citeauthoryear{{Weinberger}, {Haehnelt}  \&
  {Kulkarni}}{{Weinberger} et~al.}{2019}]{Weinberger2019}
{Weinberger} L.~H.,  {Haehnelt} M.~G.,   {Kulkarni} G.,  2019, \mn@doi [\mnras]
  {10.1093/mnras/stz481}, \href
  {https://ui.adsabs.harvard.edu/abs/2019MNRAS.485.1350W} {485, 1350}

\makeatother
\end{thebibliography}


\appendix

\section{Evolution of IGM properties with HI ionization fraction}
\label{appendix:pro_vi}
\begin{figure}
\centering
	\includegraphics[width=0.92\linewidth]{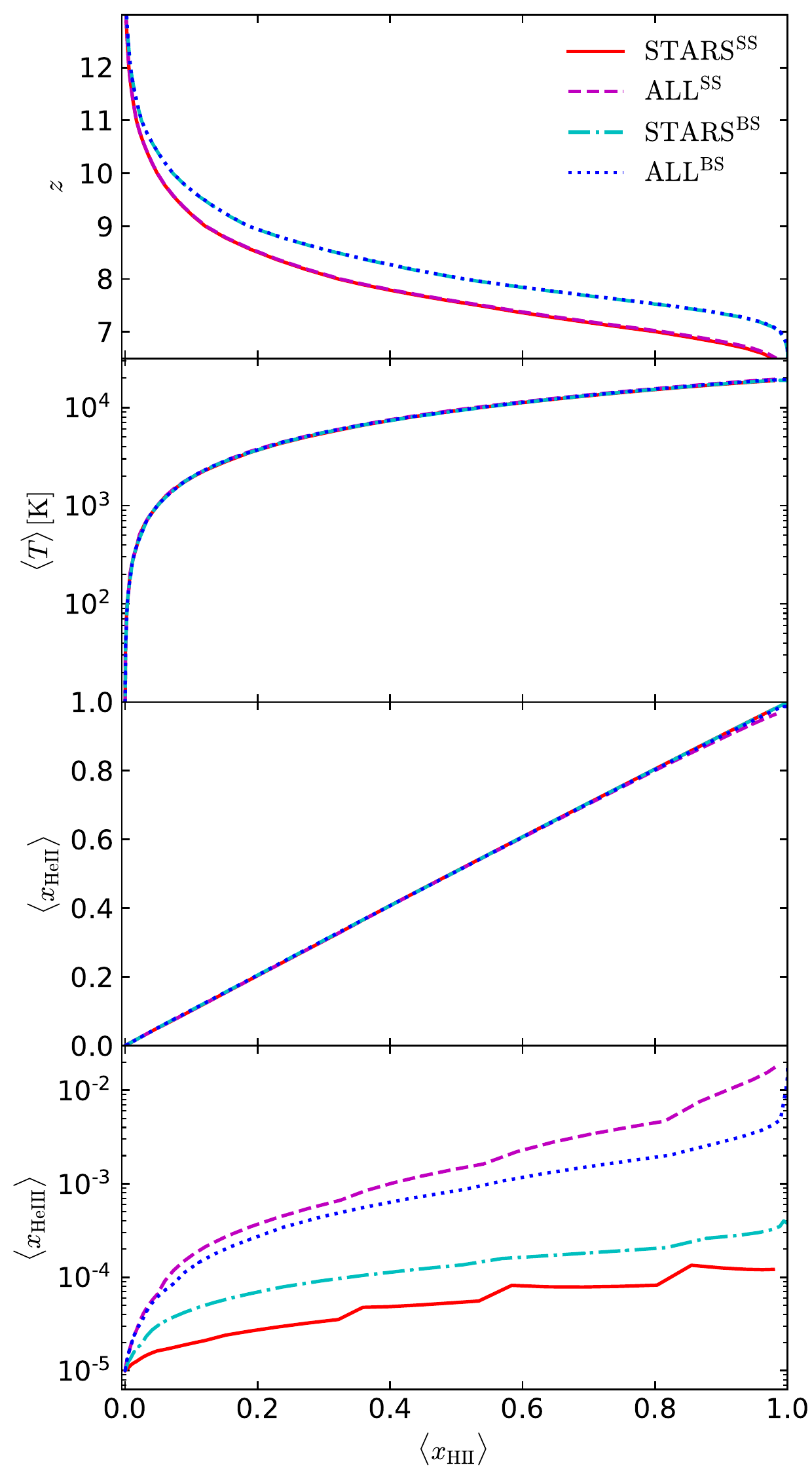}
    \caption{From top to bottom, the panels show the redshift $z$ corresponding to $\langle x_{\rm HII}\rangle$, and the evolution of the volume averaged gas temperature $\langle T \rangle$, and fractions of HeII $\langle x_{\rm HeII} \rangle$ and HeIII $\langle x_{\rm HeIII} \rangle$, as functions of $\langle x_{\rm HII}\rangle$, for simulations $\mathrm{STARS^{SS}}$ (red solid lines),  $\mathrm{ALL^{SS}}$ (magenta dashed), $\mathrm{STARS^{BS}}$ (cyan dash-dotted) and $\mathrm{ALL^{BS}}$ (blue dotted). 
    }
    \label{fig:averaged_vi}
\end{figure}

\begin{figure*}
\centering
	\includegraphics[width=0.92\linewidth]{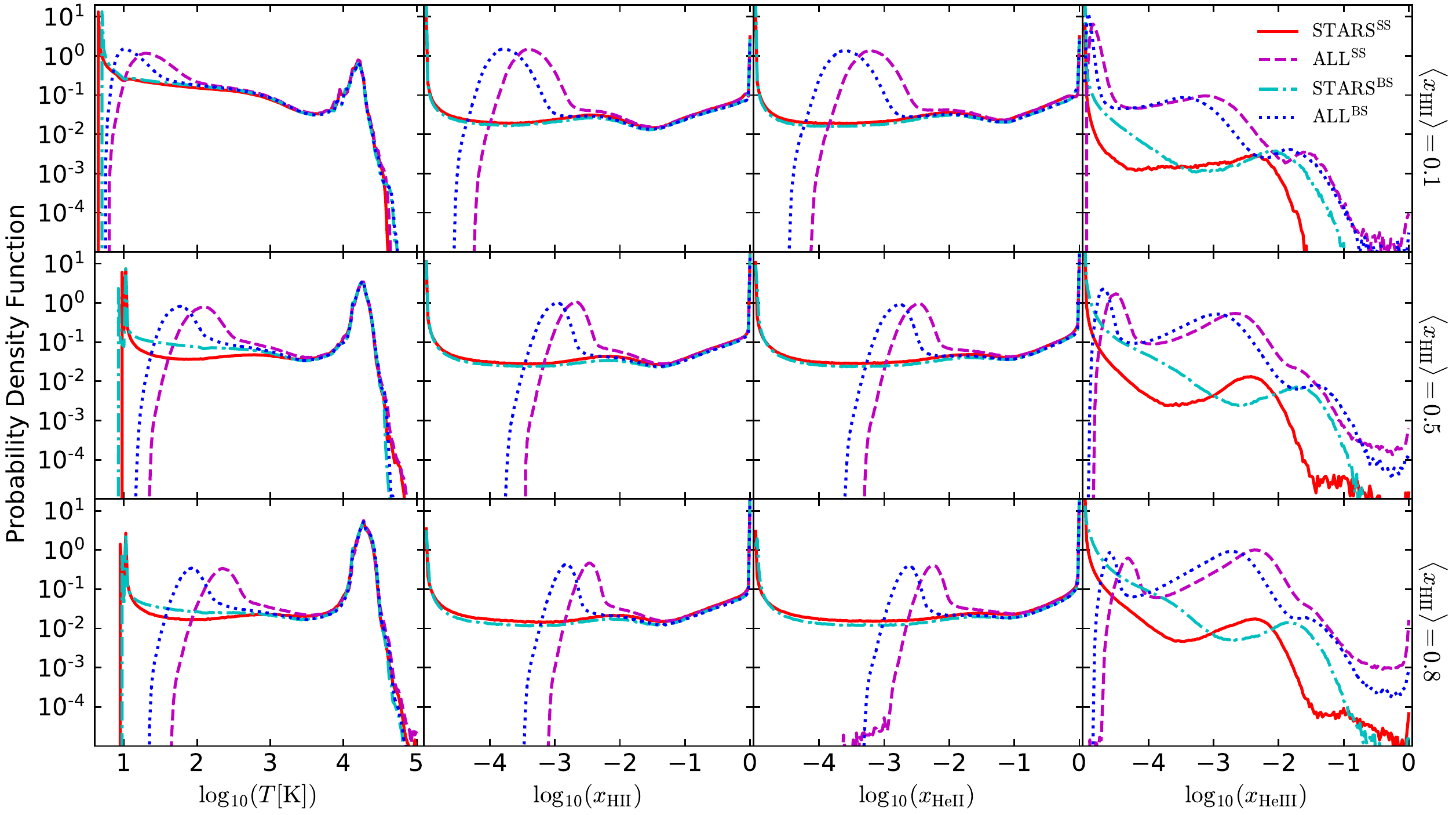}
    \caption{From left to right, probability density functions of gas temperature $T$, HII fraction $x_{\rm HII}$, HeII fraction $x_{\rm HeII}$ and HeIII fraction $x_{\rm HeIII}$ of cells from simulations $\mathrm{STARS^{SS}}$ (red solid lines), $\mathrm{ALL^{SS}}$ (magenta dashed), $\mathrm{STARS^{BS}}$ (cyan dash-dotted) and $\mathrm{ALL^{BS}}$ (blue dotted). 
    From top to bottom the rows refer to $\langle x_{\rm HII}\rangle$ = 0.1, 0.5 and 0.8. 
    }
    \label{fig:frac1_vi}
\end{figure*}

As a supplement to Figs.~\ref{fig:averaged} and~\ref{fig:frac1}, to highlight some features emerging in our model, in Figs.~\ref{fig:averaged_vi} and~\ref{fig:frac1_vi} we show the evolution of IGM properties as a function of the HII fraction.

For a constant $\langle x_{\rm HII}\rangle$, the four simulations show almost the same volume averaged gas temperature $\langle T \rangle$, i.e. the global gas temperature is dominated by the ionized gas.
It is clear that $\langle x_{\rm HeII} \rangle$ closely follows $\langle x_{\rm HII}\rangle$, although it is slightly lower than the latter at the end of the EoR.
Instead, $\langle x_{\rm HeIII} \rangle$ displays differences larger than those shown in Fig.~\ref{fig:averaged}. 
With the same $\langle x_{\rm HII}\rangle$ (here 0.1, 0.5 and 0.8), the 1-D PDFs of IGM temperature are only obviously different at $T<10^{3}\,\rm K$ (i.e. the temperature of neutral or partially ionized gas).
Simulation $\mathrm{ALL^{BS}}$ has a gas temperature at $T<10^{3}\,\rm K$ lower than that of $\mathrm{ALL^{SS}}$, due to it reaching the given ionization fraction earlier and thus having experienced shorter heating by the energetic sources.
For the same reason, similar features also appear on the 1-D PDFs of $x_{\rm HII}$ and $x_{\rm HeII}$ at $<10^{-2}$.
Similar to those in Fig.~\ref{fig:frac1}, the 1-D PDFs of $x_{\rm HeIII}$ display abundant differences among the four simulations. 
The comparison to Fig.~\ref{fig:frac1} allows one to disentangle the effects of energetic sources and stellar sources, as the former have the same total emission at a given redshift and the latter at a given ionization fraction. 
Thus, identical features at a given redshift are driven by the energetic sources (e.g. the floor of partial ionization of HII and HeII or the high $x_\mathrm{HeIII}$ tail at $z=6.5$), while those identical at given $\langle x_{\rm HII}\rangle$ are clearly caused by the stellar sources (e.g. intermediate ionization above $x_{\rm HII}=10^{-1.5}$).

\bsp	
\label{lastpage}
\end{document}